\documentclass[english,tightenlines,eqsecnum,amsmath,amssymb,aps,prd,superscriptaddress,nofootinbib,showpacs,floats]{article}
\usepackage{amsmath,amsfonts,amssymb,multicol}
\usepackage{graphicx,caption,subcaption}
\usepackage[usenames]{xcolor}
\definecolor{AHZ}{rgb}{0.0,0.9,0.2}
\usepackage[colorlinks,linkcolor=AHZ,citecolor=red,bookmarks]{hyperref}
\usepackage{rotating}
\usepackage[mathscr]{eucal}
\usepackage{graphicx}
\usepackage[T1]{fontenc}
\usepackage{babel}
\usepackage{authblk}
\usepackage{epsfig}
\usepackage{color}
\usepackage{amssymb}
\usepackage{fancyhdr}

\def\nn{\nonumber\\}
\newcommand{\f}[2]{\frac{#1}{#2}}
\def\be{\begin{equation}}
\def\ee{\end{equation}}
\def\bea{\begin{eqnarray}}
\def\eea{\end{eqnarray}}

\def\bwt{\begin{widetext}}
\def\ewt{\end{widetext}}

\usepackage{amsmath}

\usepackage{amssymb}

\begin{document}

\title{Static Vacuum Solutions on Curved Spacetimes with Torsion}
\author[1]{Hamid Shabani,\thanks{h.shabani@phys.usb.ac.ir}}
\author[2]{Amir Hadi Ziaie,\thanks{ah.ziaie@gmail.com}}
\affil[1]{Physics Department, Faculty of Sciences, University of Sistan and Baluchestan, Zahedan, Iran}
\affil[2]{Department of Physics, Kahnooj Branch, Islamic Azad University, Kerman, Iran}
\renewcommand\Authands{ and }
\maketitle
\begin{abstract}
The Einstein-Cartan-Kibble-Sciama ({\sf ECKS}) theory of gravity naturally extends Einstein\rq{}s general relativity ({\sf GR}) to include intrinsic angular momentum (spin) of matter. The main feature of this theory consists of an algebraic relation between spacetime torsion and spin of matter which indeed deprives the torsion of its dynamical content. The Lagrangian of {\sf ECKS} gravity is proportional to the Ricci curvature scalar constructed out of a general affine connection so that owing to the influence of matter energy-momentum and spin, curvature and torsion are produced and interact only through the spacetime metric. In the absence of spin, the spacetime torsion vanishes and the theory reduces to {\sf GR}. It is however possible to have torsion propagation in vacuum by resorting to a model endowed with a non-minimal coupling between
curvature and torsion. In the present work we try to investigate possible effects of the higher order terms that can be constructed from spacetime curvature and torsion, as the two basic constituents of Riemann-Cartan geometry. We consider Lagrangians that include fourth-order scalar invariants from curvature and torsion and then investigate the resulted field equations. The solutions that we find show that there could exist, even in vacuum, nontrivial static spacetimes that admit both black holes and naked singularities. 
\end{abstract}
\section{Introduction}
General relativity is the most satisfactory and accurate gravitational
theory. It provides a well-founded and consistent description of all known gravitational
phenomena. This theory describes, in a remarkable way, the gravitational interaction as a purely
geometrical effect of the spacetime so that a large number of experimental evidences
has confirmed the validity of this theory (on macroscopic scales) as the fundamental theoretical setting for modern astrophysics and cosmology \cite{grtest}. From tests of gravity performed in the solar system (weak field limit), precession of perihelia of planets, deflection of light during a solar eclipse, gravitational redshift and lensing effects to observations of double neutron stars, and recent discovery of gravitational waves from a binary black hole merger \cite{gwave}, {\sf GR} has passed all observational tests \cite{obsertest,obsertest1}. However, this has not stopped a various of efforts to build a better gravitation theory, at least at the classical level we might hope to obtain a richer theory being in closer agreement with experiments. In this sense, physically viable ideas to extend {\sf GR} have always been welcome owing to the fundamental concepts deeply rooted within physics behind the observed phenomena and open problems which are still unsolved by the theory. Among such question still pending to be solved one may quote from the understanding of astrophysical as well as cosmological spacetime singularities or the nature of dark matter ({\sf DM}), dark energy ({\sf DE}) or inflationary era in the very early Universe \cite{infdedm,infdedm1,infdedm2,infdedm3}. Moreover, it could also be remarkable that an improved classical theory of {\sf GR} will lead to the much-dreamed-of renormalizable quantum theory of gravity. On the other hand, in the regimes where strong gravitational fields are present, e.g., in the evolution of universe at very early times, regions with extreme curvature around the black holes and neutron stars, little progress has been made in examining the predictions of {\sf GR} \cite{strbhneugr}. There are two reasons for this mustiness. At first sight, phenomena that occur in the regimes with strong gravitational fields are complicated and are often explosively progressed, making it very tough to detect observable properties that depend purely on the gravitational field and that allow for quantitative tests of gravity theories. The second problem stems from the non-existence of general theoretical framework within which quantified deviations from the predictions of {\sf GR} in the strong field regime could be recognized. \par

Our current comprehension of the universe leaves very little skepticism that {\sf GR} itself fails to predict the physics behind the phenomena occurring at the limit of very strong gravitational fields. For example, the study of Damour and Esposito-Farese \cite{DaEsFa} revealed one of the main reasons that necessitate careful theoretical studies of possible extensions of {\sf GR} that could possibly provide a befitting framework with a unique precision strong-field test of gravity. Another important issue deals with providing correctly a basis for including the angular momentum of gravitating sources in the presence of a dynamical
spacetime within the same framework. Of most interest is the inclusion of intrinsic angular momentum of matter (spin) within the {\sf GR} theory. As we know, all elementary particles can be classified by means of irreducible unitary representations of Poincare group and can be labeled by mass and their spin, so that, the former is related to the translational part of the Poincare group and the latter with the rotational part. Therefore, a matter distribution with energy-momentum and spin over the spacetime leads to the field theoretical notions of an energy momentum tensor ({\sf EMT}), ${\sf T}_{\mu\nu}$, and spin angular momentum tensor $\tau^{\alpha\beta\gamma}$ of matter\footnote{The work of Cartan was largely forgotten until
the 1940s when Papapetrou \cite{PAPSPIn} and Weyssenhoff and Raabe \cite{WEYRABB} showed that the energy momentum tensor of a spinning particle is asymmetric. Since the source
of gravity in {\sf GR} is a symmetric {\sf EMT}, it is clear that an extension of the theory is necessary for matter distributions with intrinsic spin.} \cite{hehlvonkerlick}. In the macroscopic limit, mass (or correspondingly energy-momentum) adds up due to its monopole character while, the spin which is of dipole character usually averages out. Owing to vanishing effects of spin at macroscopic limit, the dynamical characterization of a continuous matter can be successfully achieved by energy-momentum alone. As we have learned from {\sf GR}, the {\sf EMT} of matter is the source of gravitational field which is coupled to spacetime metric ${\sf g}_{\mu\nu}$ through a Riemannian manifold. However, when we penetrate into the microscopical realm of matter, we figure out that the spin of matter also comes into play and determines the dynamics of matter distribution. Hence, we expect that, in analogy to the coupling of energy-momentum to the spacetime metric, spin is coupled to a geometrical quantity, i.e., the spacetime torsion, and therefore acts as a source of gravitational field. However, the Riemannian spacetime of {\sf GR} could not provide a suitable setting in order to introduce the spacetime torsion and has to be slightly generalized to the Riemann-Cartan spacetime ${\sf U}_4$. In this regard, it sounds relevant to look upon {\sf GR} as the macroscopic limit of a still illustrative  microphysical gravity theory. The first attempts towards reaching the simplest generalization of {\sf GR} can be traced back to Elie Cartan\rq{}s papers \cite{ecartan,ecartan1,ecartan2,ecartan3} who suggested that spacetime torsion could be exploited as the macroscopic manifestation of the intrinsic angular momentum of matter. After this idea, Kibble and Sciama established the foundations of ${\sf U}_4$ theory and highlighted the geometrical role of spacetime torsion in modern physics \cite{KibScia,KibScia1,KibScia2,KibScia3,KibScia4,KibScia5,KibScia6}.\par
Today, the {\sf ECKS} gravity is known as a viable description of the gravitational field that takes into account the spin degrees of freedom of matter and describes their influence on the geometrical structure of spacetime. The latter is carried by non-trivial curvature and torsion, and usually called the Riemann-Cartan spacetime ${\sf U}_4$. By coupling the energy-momentum density and the intrinsic angular momentum of the matter to the metric and the torsion tensors respectively and by treating them as independent fields, the {\sf ECKS} gravity provides the simplest classical generalization of
{\sf GR}. The predictions of the theory differ from those of
{\sf GR} only at extremely high densities, like those expected in the interior of black holes and extremely high matter densities that may have been present in the early stages of the universe \cite{ecksbheui,Venzo-Hehl,Venzo-Hehl1,Venzo-Hehl2,Venzo-Hehl3,Venzo-Hehl4,Gasper}. In such superdense regimes, the coupling between spin of matter and spacetime torsion leads to a repulsive gravitational \lq\lq{}force\rq\rq{},
which could (in principle at least) prevent the formation of spacetime singularities \cite{gasperiniavoid,gasperiniavoid1,gasperiniavoid10,gasperiniavoid100,gasperiniavoid2}.
\par
Since its reemergence in the late 1950s there have been renewed interests in {\sf ECKS} theory in recent years. As the theory is still considered viable and remains an active field of research, much attempt have been made to generalize {\sf ECKS} gravity in order to incorporate torsion into novel quantum theories and therefore providing possible extensions of {\sf GR} to theories of micro-physical interactions \cite{torquantum,torquantum1,torquantum2,torquantum3}, exploring cosmological implications of torsion~\cite{Putzfield}, emergent universe scenario \cite{emerun}, gravitational collapse \cite{grcolapse,grcolapse1} and higher dimensional gravity theories \cite{highertor,highertor1,highertor2}. Work along this line has been carried out in black hole physics \cite{bhtorsion,bhtorsion1,bhtorsion2,bhtorsion3,bhtorsion4,bhtorsion5} where possible effects of torsion can be investigated from astrophysical viewpoint. The importance of black holes for gravitational physics is obvious as their existence can be used as a testbed towards our understanding of strong gravitational fields, beyond the point of small corrections to Newtonian physics and a test of our understanding of astrophysics, particularly of stellar evolution \cite{strongfieldbh}. In this respect, it could be interesting to investigate physics of black holes in the context of gravitational theories with more degrees of freedom e.g. {\sf ECKS} gravity. However, in this theory, the torsion field equation is unusual as the relationship between spin of matter and spacetime torsion is a pure algebraic equation and not differential. This means the torsion field does not propagate through spacetime as some other fields do. In this form the theory lends itself to straightforward comparison with {\sf GR}. Though in {\sf ECKS} theory, the spacetime torsion can not propagate outside of the matter distribution, it is possible to build Lagrangians with higher order corrections describing a dynamical torsion field \cite{highlagdyntor}. Based on these considerations, it may be possible that the spacetime torsion and curvature couple to each other and have direct interaction in the regimes of extreme gravity which are present in a black hole spacetime. Motivated by this idea, our aim in the herein work is to study consequences of including higher order terms from spacetime curvature and torsion and to obtain exact static vacuum spacetimes. The lay out of the present work is as follows. In section \ref{I} we try to construct the corresponding Lagrangian describing interaction between spacetime torsion and curvature. The field equations are then found using variation with respect to torsion and metric, independently. The static vacuum solutions to the field equations are found in section \ref{II} and related discussion about these solution are given, correspondingly. Finally, the conclusions of our study are given in section \ref{conclu}.
\section{Modified action and field equations}\label{I}
As we know, the dynamics of the gravitational field, i.e., the metric field, in {\sf GR} is described by the Hilbert-Einstein action with the Lagrangian which is linear in curvature scalar. Contrary to {\sf GR}, in the gravity with torsion there is a considerable freedom in constructing the dynamical scheme, since one can define much more invariants from torsion and curvature tensors. Two most attractable classes of models, namely, the {\sf ECKS} theory and the quadratic theories. Let us begin with this  theory in which the spacetime torsion is introduced as the antisymmetric part of the general affine
connection
\be\label{TEQ}
{\sf Q}^{\alpha}_{\,\,\mu\nu}=\tilde{\Gamma}^{\alpha}_{~\mu\nu}-\tilde{\Gamma}^{\alpha}_{~\nu\mu},~~~~{\sf Q}^{\alpha}_{\,\,\mu\nu}=-{\sf Q}^{\alpha}_{\,\,\nu\mu},
\ee
The Riemann curvature tensor constructed by these connections is given as
\bea\label{ECRI}
\tilde{{\sf R}}^{\lambda}_{\,\,\mu\nu\rho}=\partial_{\nu}\tilde{\Gamma}^{\lambda}_{\,\,\mu\rho}-
\partial_{\rho}\tilde{\Gamma}^{\lambda}_{\,\,\mu\nu}+\tilde{\Gamma}^{\sigma}_{\,\,\mu\rho}
\tilde{\Gamma}^{\lambda}_{\,\,\sigma\nu}-\tilde{\Gamma}^{\sigma}_{\,\,\mu\nu}
\tilde{\Gamma}^{\lambda}_{\,\,\sigma\rho}.
\eea
The affine connection is metric compatible, i.e., $\tilde{\nabla}_{\alpha}\textsf{g}_{\mu\nu}=0$ and thus can be written as
\be\label{GAFC}
\tilde{\Gamma}^{\alpha}_{\,\,\beta\gamma}=\Gamma^{\alpha}_{\,\,\beta\gamma}+\textsf{K}^{\alpha}_{\,\,\beta\gamma},
\ee
where $\Gamma^{\alpha}_{\,\,\beta\gamma}$ is the Levi-Civita Christoffel connection and
\be\label{contt}
\textsf{K}^{\alpha}_{\,\,\,\beta\gamma}=\f{1}{2}\left[{\sf Q}^\alpha\!\!_{\beta\gamma}-{\sf Q}_{\beta~\gamma}^{\,\,\,\alpha}-\textsf{Q}_{\gamma~\beta}^{\,\,\alpha}\right],~~~~{\sf K}_{\alpha\beta\gamma}=-{\sf K}_{\beta\alpha\gamma},
\ee
is defined as the contorsion tensor.
Substituting the above expression into (\ref{ECRI}) we get
\bea\label{ECRI1}
\tilde{{\sf R}}^{\lambda}_{\,\,\,\mu\nu\rho}&=&{\sf R}^{\lambda}_{~\mu\nu\rho}+{\sf K}^{\lambda}_{~\mu\rho,\nu}-{\sf K}^{\lambda}_{~\mu\nu,\rho}\nn
&+&\Gamma^{\sigma}_{\,\,\mu\rho}{\sf K}^{\lambda}_{~\sigma\nu}+\Gamma^{\lambda}_{\,\,\sigma\nu}{\sf K}^{\sigma}_{~\mu\rho}+{\sf K}^{\sigma}_{~\mu\rho}{\sf K}^{\lambda}_{~\sigma\nu}\nn
&-&\Gamma^{\sigma}_{\,\,\mu\nu}{\sf K}^{\lambda}_{~\sigma\rho}-\Gamma^{\lambda}_{\,\,\sigma\rho}{\sf K}^{\sigma}_{~\mu\nu}-{\sf K}^{\sigma}_{~\mu\nu}{\sf K}^{\lambda}_{~\sigma\rho},\nn
\eea
where ${\sf R}^{\lambda}_{\,\,\,\mu\nu\rho}$ denotes Riemann curvature tensor constructed out of Christoffel connection. Contracting twice gives the Ricci curvature scalar as
\bea\label{RICCI}
\tilde{\sf R}&=&{\sf R}+{\sf g}^{\mu\rho}{\sf K}^{\nu}\!\!~_{\mu\rho,\nu}-{\sf g}^{\mu\rho}{\sf K}^{\nu}\!\!~_{\mu\nu,\rho}+{\sf g}^{\mu\rho}\Gamma^{\sigma}_{\,\,\mu\rho}{\sf K}^{\nu}\!\!~_{\sigma\nu}\nn
&-&{\sf g}^{\mu\rho}\Gamma^{\sigma}_{\,\,\mu\nu}{\sf K}^{\nu}\!\!~_{\sigma\rho}-{\sf g}^{\mu\rho}\Gamma^{\nu}_{\,\,\sigma\rho}{\sf K}^{\sigma}\!\!~_{\mu\nu}-{\sf g}^{\mu\rho}{\sf K}^{\sigma}\!\!~_{\mu\nu}{\sf K}^{\nu}\!\!~_{\sigma\rho}\nn
&+&\Gamma^{\lambda}_{\,\,\sigma\lambda}{\sf K}^{\sigma\mu}\!\!~_{\mu}+{\sf K}^{\sigma\mu}\!\!~_{\mu}{\sf K}^{\lambda}\!\!~_{\sigma\lambda}.
\eea
The metricity condition in {\sf GR} which is defined as, ${\nabla}_{\alpha}{\sf g}_{\mu\nu}=0$, leaves us with the following expressions
\bea\label{PMET}
{\sf g}^{\mu\rho}_{~~,\nu}&=&-{\sf g}^{\alpha\rho}\Gamma^{\mu}_{\,\,\nu\alpha}-{\sf g}^{\mu\alpha}\Gamma^{\rho}_{\,\,\nu\alpha},\nn
{\sf g}^{\mu\rho}_{~~,\rho}&=&-{\sf g}^{\alpha\rho}\Gamma^{\mu}_{\,\,\rho\alpha}-{\sf g}^{\mu\alpha}\Gamma^{\rho}_{\,\,\rho\alpha},
\eea
by the virtue of which we can simplify (\ref{RICCI}) to finally get
\bea\label{RICCIF}
\tilde{{\sf R}}&=&{\sf R}-2{\sf K}^{\lambda\beta}\!\!~_{\lambda;\beta}+{\sf K}^{\gamma\mu}\!\!~_{\mu}{\sf K}^{\lambda}\!\!~_{\gamma\lambda}-{\sf K}^{\sigma\rho\nu}{\sf K}_{\nu\sigma\rho}\nn
&=&{\sf R}+\f{1}{4}{\sf Q}_{\beta\gamma\lambda}{\sf Q}^{\beta\gamma\lambda}+\f{1}{2}{\sf Q}^{\beta\gamma\lambda}{\sf Q}_{\gamma\beta\lambda}+{\sf Q}^{\beta\,\,\gamma}_{\,\,\beta}{\sf Q}^{\lambda}_{\,\,\gamma\lambda}+2{\sf Q}^{\beta\,\,\gamma}_{\,\,\beta\,\,\,;\gamma},
\eea
where $;\equiv\nabla$ is the covariant derivate constructed from the Christoffel connection. We therefore obtain the gravitational action for {\sf ECKS} theory as
\bea\label{ECaction}
{\sf S}&=&\int d^4x\sqrt{-{\sf g}}\left[-\f{\tilde{{\sf R}}}{\kappa^2}+\mathcal{L}_m\right]\nn
&=&\int d^4x\sqrt{-{\sf g}}\bigg\{\f{-1}{\kappa^2}\bigg[{\sf R}+\f{1}{4}{\sf Q}_{\beta\gamma\lambda}{\sf Q}^{\beta\gamma\lambda}+\f{1}{2}{\sf Q}^{\beta\gamma\lambda}{\sf Q}_{\gamma\beta\lambda}\nn&+&{\sf Q}^{\beta\,\,\gamma}_{\,\,\beta}{\sf Q}^{\lambda}_{\,\,\gamma\lambda}+2{\sf Q}^{\beta\,\,\gamma}_{\,\,\beta\,\,\,;\gamma}\bigg]+\mathcal{L}_m\bigg\},\nn
\eea
where $\kappa^2=\f{8\pi G}{c^4}$ and $\mathcal{L}_m=\mathcal{L}_m\left(\Psi;\Psi_\alpha;{\sf g}_{\alpha\beta};{\sf g}_{\alpha\beta,\gamma};{\sf Q}^{\alpha}_{\,\,\,\,\beta\gamma}\right)$ being the gravitational coupling constant and Lagrangian of minimally coupled matter field(s) $\Psi$.  We are interested here to consider vacuum solutions therefore the {\sf ECKS} action is not suitable for such a purpose since in this theory the torsion vanishes in the absence of matter. We then consider the following action constructed out of scalar invariants of spacetime curvature and torsion which is given as
\bea\label{ECmod1}
{\sf S}&\!\!\!\!\!=\!\!\!\!\!\!\!&\int\!\! d^4x\sqrt{-{\sf g}}\Bigg\{\!\!-\f{1}{\kappa^2}\Bigg[\tilde{{\sf R}}+\tilde{{\sf R}}\bigg({\sf Q}_{\alpha\beta\gamma}\left[a_{1}{\sf Q}^{\alpha\beta\gamma}+a_{2}{\sf Q}^{\beta\alpha\gamma}\right]+a_{3}{\sf Q}^{\alpha\,\,\beta}_{\,\,\,\alpha} {\sf Q}^{\gamma}_{\,\,\beta\gamma}\bigg)\Bigg]+\mathcal{L}_m\Bigg\},\nn
\eea
where $a_1-a_{3}$ are coupling constants. 
 Using the expression (\ref{RICCIF}) for Ricci scalar we have
\bea\label{ECactionmodified}
{\sf S}=\int d^4x\sqrt{-{\sf g}}\Bigg\{-\f{1}{\kappa^2}\Big(\mathbb{L}+\sum_{i=1}^{3} a_{i}{\mathfrak L}_{i} \Big)+\mathcal{L}_m\Bigg\},
\eea
where the {\sf ECKS} Lagrangian $\mathbb{L}$ and $\mathfrak{L}_{1}-\mathfrak{L}_{3}$ are defined as
\bea\label{act-Ec}
\mathbb{L}={\sf R}+\f{1}{4}{\sf Q}_{\alpha\beta\gamma}{\sf Q}^{\alpha\beta\gamma}+\f{1}{2}{\sf Q}^{\alpha\beta\gamma}{\sf Q}_{\beta\alpha\gamma}+{\sf Q}_{\,\,\alpha}^{\alpha\,\beta}{\sf Q}^{\gamma}_{\,\,\beta\gamma}+2 {\sf Q}^{\alpha\,\,\beta}_{\,\,\alpha\,\, ;\beta},
\eea

\bea\label{act-1}
{\mathfrak L}_{1}&=&{\sf R}{\sf Q}_{\alpha\beta\gamma}{\sf Q}^{\alpha\beta\gamma}+\f{1}
{4}{\sf Q}_{\alpha\beta\gamma}{\sf Q}^{\alpha\beta\gamma}\Big{[}{\sf Q}^{\delta\epsilon\eta}\big(2{\sf Q}_{\epsilon\delta\eta}+{\sf Q}_{\delta\epsilon\eta}\big)+8{\sf Q}^{\delta\,
\,\,\epsilon}_{\,\,\delta\,\,;\epsilon}\Big]\nn&+&{\sf Q}_{\,\,\alpha}^{\alpha\,\beta}{\sf Q}^{\gamma}_{\,\,
\beta\gamma}{\sf Q}^{\delta\epsilon\eta}{\sf Q}_{\delta\epsilon\eta},
\eea

\bea\label{act-2}
{\mathfrak L}_{2}&=&{\sf R}{\sf Q}_{\alpha\beta\gamma}{\sf Q}^{\beta\alpha\gamma}+\f{1}
{4}{\sf Q}_{\alpha\beta\gamma}{\sf Q}^{\alpha\beta\gamma}{\sf Q}^{\delta\epsilon\eta}{\sf Q}_{\epsilon\delta\eta}\nn&+&\f{1}
{2}{\sf Q}^{\alpha\beta\gamma}{\sf Q}_{\beta\alpha\gamma}\Big({\sf Q}_{\epsilon\delta\eta}{\sf Q}^{\delta\epsilon\eta}+4 
{\sf Q}^{\delta\,\,\epsilon}_{\,\,\delta\,\,;\epsilon}\Big)+{\sf Q}_{\,\,\alpha}^{\alpha\,\beta}Q^{\gamma}_{\,\,
\beta\gamma}{\sf Q}^{\delta\epsilon\eta}{\sf Q}_{\epsilon\delta\eta},
\eea

\bea\label{act-3}
{\mathfrak L}_{3}&=&{\sf R}{\sf Q}^{\alpha\,\,\beta}_{\,\,\,\alpha}{\sf Q}^{\gamma}_{\,\,\,\,\beta\gamma}+{\sf Q}_{\,\,\alpha}^{\alpha\,\beta}\Big(\f{1}{4}{\sf Q}^{\gamma}_{\,\,
\beta\gamma}{\sf Q}^{\delta\epsilon\eta}{\sf Q}_{\delta\epsilon\eta}+\f{1}{2}{\sf Q}^{\gamma}_{\,\,
\beta\gamma}{\sf Q}^{\delta\epsilon\eta}{\sf Q}_{\epsilon\delta\eta}\nn&+&{\sf Q}^{\eta}_{\,\,\epsilon\eta}{\sf Q}^{\gamma}_{\,\,
\beta\gamma}{\sf Q}^{\delta\,\,\,\epsilon}_{\,\,\delta}+2{\sf Q}^{\gamma}_{\,\,\beta\gamma}{\sf Q}^{\delta\,\,\,\epsilon}_{\,\,\,
\delta\,\,;\epsilon}\Big),
\eea

In the present work we take the coupling constant $a_1$ to be nonzero in action (\ref{ECactionmodified}) and set the rest of constants to be vanished. We here have two independent tensor fields, i.e., the metric and torsion fields and thus we expect individual field equations for each of them. Consequently, varying action (\ref{ECactionmodified}) with respect to the torsion field ${\sf Q}_{\alpha\beta\gamma}$, leaves us with the following equation of motion for spacetime torsion as
\bea\label{eqmtor-1}
\mathbb{E}^{ \alpha\beta\gamma}+a_{1}\mathfrak{E}^{\alpha\beta\gamma}=\kappa^2\tau^{\alpha\beta\gamma},
\eea
where
\bea\label{eqmtor-2}
\mathbb{E}^{ \alpha\beta\gamma}=\f{1}{2}{\sf Q}^{\alpha\beta\gamma}+\f{1}{2}{\sf Q}^{\beta\alpha\gamma}-\f{1}{2}{\sf Q}^{\gamma\alpha\beta}-{\sf g}^{\alpha\gamma}{\sf Q}^{\delta\beta}_{\,\,\,\,\,\delta}+{\sf g}^{\alpha\beta}{\sf Q}^{\delta\gamma}_{\,\,\,\,\,\delta},
\eea
and
\bea\label{eqmtor-3}
\mathfrak{E}^{\alpha\beta\gamma}&=&{\sf Q}^{\alpha\beta\gamma}\Bigg[2{\sf R}+{\sf Q}^{\delta\epsilon\zeta}\big({\sf Q}_{\delta\epsilon\zeta}+{\sf Q}_{\epsilon\delta\zeta}\big)+2{\sf Q}^{\delta\,\,\epsilon}_{\,\,\delta}{\sf Q}^{\zeta}_{\,\,\epsilon\zeta}\Bigg]\nn&+&\hspace{-0.2cm}\f{1}{2}{\sf Q}_{\delta\epsilon\zeta}{\sf Q}^{\delta\epsilon\zeta}\bigg({\sf Q}^{\beta\alpha\gamma}-{\sf Q}^{\gamma\alpha\beta}\bigg)+{\sf Q}_{\epsilon\zeta\eta}{\sf Q}^{\epsilon\zeta\eta}\left[{\sf g}^{\alpha\beta}{\sf Q}^{\delta\gamma}_{\,\,\,\,\,\delta}-{\sf g}^{\alpha\gamma}{\sf Q}^{\delta\beta}_{\,\,\,\,\,\delta}\right]\nn&+&\hspace{-0.2cm}2{\sf Q}^{\delta\epsilon\zeta}\left[{\sf g}^{\alpha\gamma}{\sf Q}_{\delta\epsilon\zeta}^{\,\,\,\,\,\,;\beta}-{\sf g}^{\alpha\beta}{\sf Q}_{\delta\epsilon\zeta}^{\,\,\,\,\,\,;\gamma}\right]+4{\sf Q}^{\alpha\beta\gamma}{\sf Q}^{\delta\,\,\,\epsilon}_{\,\,\delta\,\,;\epsilon},
\eea
where $\tau^{\alpha\beta\gamma}$ is defined as the spin angular momentum tensor given by
\be\label{spinanmom}
\tau^{\alpha\beta\gamma}=\f{1}{\sqrt{-{\sf g}}}\f{\delta\left[\sqrt{-{\sf g}}\mathcal{L}_m\right]}{\delta {\sf Q}_{\alpha\beta\gamma}}.
\ee
Varying action (\ref{ECactionmodified}) with respect to the metric field we get 
\bea\label{eqmet-1}
\mathcal{\mathbb{G}}_{ \mu\nu}+a_{1}{\mathfrak G}_{ \mu\nu}=\kappa^2{\sf T}_{ \mu\nu},
\eea
where
\bea\label{eqmet-2}
\mathcal{\mathbb{G}}_{ \mu\nu}={\sf G}_{ \mu\nu}&-&\f{1}{4}g_{\mu\nu}\left[{\sf Q}^{\alpha\beta\gamma}{\sf Q}_{\beta\alpha\gamma}+\f{1}{2}{\sf Q}_{\alpha\beta\gamma}{\sf Q}^{\alpha\beta\gamma}+2{\sf Q}^{\alpha\,\,\beta}_{\,\,\alpha}{\sf Q}^{\gamma}_{\,\,\beta\gamma}\right]\nn&+&\f{1}{2}\bigg[{\sf Q}^{\alpha\,\,\beta}_{\,\,\mu}{\sf Q}_{\alpha\nu\beta}+{\sf Q}^{\alpha\,\,\beta}_{\,\,\mu}{\sf Q}_{\beta\nu\alpha}-2{\sf Q}^{\alpha}_{\,\,\mu\alpha}{\sf Q}^{\beta}_{\,\,\nu\beta}-\f{1}{2}{\sf Q}_\mu^{\,\,\alpha\beta}{\sf Q}_{\nu\alpha\beta}\bigg],\nn
\eea
and
\bea\label{eqmet-3}
\mathfrak{G}_{\mu\nu}&=&{\sf G}_{\mu\nu}{\sf Q}_{\alpha\beta\gamma}{\sf Q}^{\alpha\beta\gamma}+2{\sf R}\left[{\sf Q}^{\alpha\,\,\beta}_{\,\,\mu}{\sf Q}_{\alpha\nu\beta}-\f{1}{2}{\sf Q}_{\mu}^{\,\,\alpha\beta}{\sf Q}_{\nu\alpha\beta}\right]\nn
&+&{\sf Q}_{\gamma\delta\epsilon}{\sf Q}^{\gamma\delta\epsilon}\bigg\{{\sf Q}^{\alpha\,\,\beta}_{\,\,\mu}\left[{\sf Q}_{\alpha\nu\beta}+\f{1}{2}{\sf Q}_{\beta\nu\alpha}\right]-{\sf Q}^{\alpha}_{\,\,\mu\alpha}{\sf Q}^{\beta}_{\,\,\nu\beta}-\f{1}{8}{\sf g}_{\mu\nu}{\sf Q}_{\alpha\beta\rho}{\sf Q}^{\alpha\beta\rho}\nn
&-&\f{1}{2}{\sf g}_{\mu\nu}{\sf Q}^{\alpha\,\,\beta}_{\,\,\alpha}{\sf Q}^{\gamma}_{\,\,\beta\gamma}-\f{1}{2}{\sf Q}_{\mu}^{\,\,\alpha\beta}{\sf Q}_{\nu\alpha\beta}\bigg\}+{\sf Q}^{\gamma\delta\epsilon}{\sf Q}_{\delta\gamma\epsilon}\bigg[{\sf Q}^{\alpha\,\,\beta}_{\,\,\mu}{\sf Q}_{\alpha\nu\beta}-\f{1}{4}{\sf g}_{\mu\nu}{\sf Q}_{\alpha\beta\rho}{\sf Q}^{\alpha\beta\rho}\nn
&-&\f{1}{2}{\sf Q}_{\mu}^{\,\,\alpha\beta}{\sf Q}_{\nu\alpha\beta}\bigg]+{\sf Q}^{\epsilon}_{\,\,\delta\epsilon}{\sf Q}^{\gamma\,\,\delta}_{\,\,\gamma}\bigg[2{\sf Q}^{\alpha\,\,\beta}_{\,\,\mu}{\sf Q}_{\alpha\nu\beta}-{\sf Q}_{\mu}^{\,\,\alpha\beta}{\sf Q}_{\nu\alpha\beta}\bigg]-2 {\sf Q}^{\alpha\beta\gamma}_{\,\,\,\,\,\,\,\,\,\,;_\mu} {\sf Q}_{\alpha\beta\gamma;_\nu}\nn
&+&{\sf g}_{\mu\nu}\bigg[2{\sf Q}^{\alpha\,\,\beta}_{\,\,\alpha}{\sf Q}^{\gamma\delta\epsilon}{\sf Q}_{\gamma\delta\epsilon;\beta}+2{\sf Q}^{\alpha\beta\gamma}\Box {\sf Q}_{\alpha\beta\gamma}+2{\sf Q}_{\alpha\beta\gamma;\delta}{\sf Q}^{\alpha\beta\gamma;\delta}\bigg]\nn&+&2\bigg[2{\sf Q}^{\alpha\,\,\beta}_{\,\,\mu}{\sf Q}_{\alpha\nu\beta}-{\sf Q}_\mu^{\,\,\alpha\beta}{\sf Q}_{\nu\alpha\beta}\bigg]{\sf Q}^{\gamma\,\,\delta}_{\,\,\,\gamma\,\,;\delta}+2\bigg[{\sf Q}^{\beta\gamma\delta}{\sf Q}^{\alpha}_{\,\,\nu\alpha}{\sf Q}_{\beta\gamma\delta;\mu}\nn
&+&{\sf Q}^{\alpha}_{\,\,\mu\alpha}{\sf Q}^{\beta\gamma\delta}{\sf Q}_{\beta\gamma\delta;\nu}\bigg]-{\sf Q}^{\alpha\beta\gamma}\bigg[{\sf Q}_{\alpha\beta\gamma;\nu;\mu}+{\sf Q}_{\alpha\beta\gamma;\mu;\nu}\bigg],
\eea
where 
\be\label{STENT}
{\sf T}_{\mu\nu}=\f{1}{\sqrt{-{\sf g}}}\f{\delta\left(\sqrt{-{\sf g}}{\mathcal L}_m\right)}{\delta {\sf g}^{\mu\nu}},~~~~~~{\sf G}_{\mu\nu}={\sf R}_{\mu\nu}-\f{1}{2}{\sf g}_{\mu\nu}{\sf R},~~~{\rm and}~~\Box\equiv\nabla_\beta\nabla^\beta,
\ee
are defined as the stress-energy tensor of matter fields, Einstein tensor and the d\rq{}alembert operator. We note that for $a_1=0$ the field equations (\ref{eqmtor-1}) and (\ref{eqmet-1}) will reduce to those of standard {\sf ECKS} theory. Therefore it can be easily seen from equation (\ref{eqmtor-1}) that for $a_1=0$ and $\tau^{\alpha\beta\gamma}=0$, the left hand side as given in (\ref{eqmtor-2}) must  be zero. Since this expression does not contain any differential of torsion, we then conclude that the spacetime torsion must vanish in vacuum. However, in contrast to {\sf ECKS} theory where the equation governing the spacetime torsion is of pure algebraic type, we here have a dynamical equation which allows the spacetime torsion to propagate even in the absence of spin of matter. Such a behavior could not be seen in {\sf ECKS} theory where the torsion vanishes outside the matter distribution \cite{Venzo-Hehl}.  In case in which $a_1=0$ the spacetime torsion can be obtained from (\ref{eqmtor-1}) via introducing a suitable spin source e. g., Weyssenhoff fluid \cite{weyspinsource,weyspinsource1,weyspinsource2,weyspinsource3,weyspinsource4,weyspinsource5}. Therefore, substituting for the torsion into equation (\ref{eqmet-1}) and after a few simplification we will get the so called combined field equations of {\sf ECKS} theory \cite{Venzo-Hehl,Venzo-Hehl1,Venzo-Hehl2,Venzo-Hehl3,Venzo-Hehl4,Gasper}. However, for $a_1\neq0$,  such a process is completely different to the usual case as one needs to solve for a set of differential equations for torsion and spin as a source to find the dynamics of spacetime torsion. Our focus then here is to search for exact spacetimes that represent static solutions in the absence of matter fields.
\section{Solutions to the field equations}\label{II}
In this section we seek for static vacuum solutions to the field equations (\ref{eqmtor-1}) and (\ref{eqmet-1}) and study their properties. The exact static solutions that we shall find show remarkable deviation from those of {\sf GR} and could provide a setting to investigate the possible effects of torsion on spacetime geometry.
\subsection{Class A Solutions}\label{IIA}
\begin{figure}[htbp]
\hspace*{-1cm}
\includegraphics[scale=0.39]{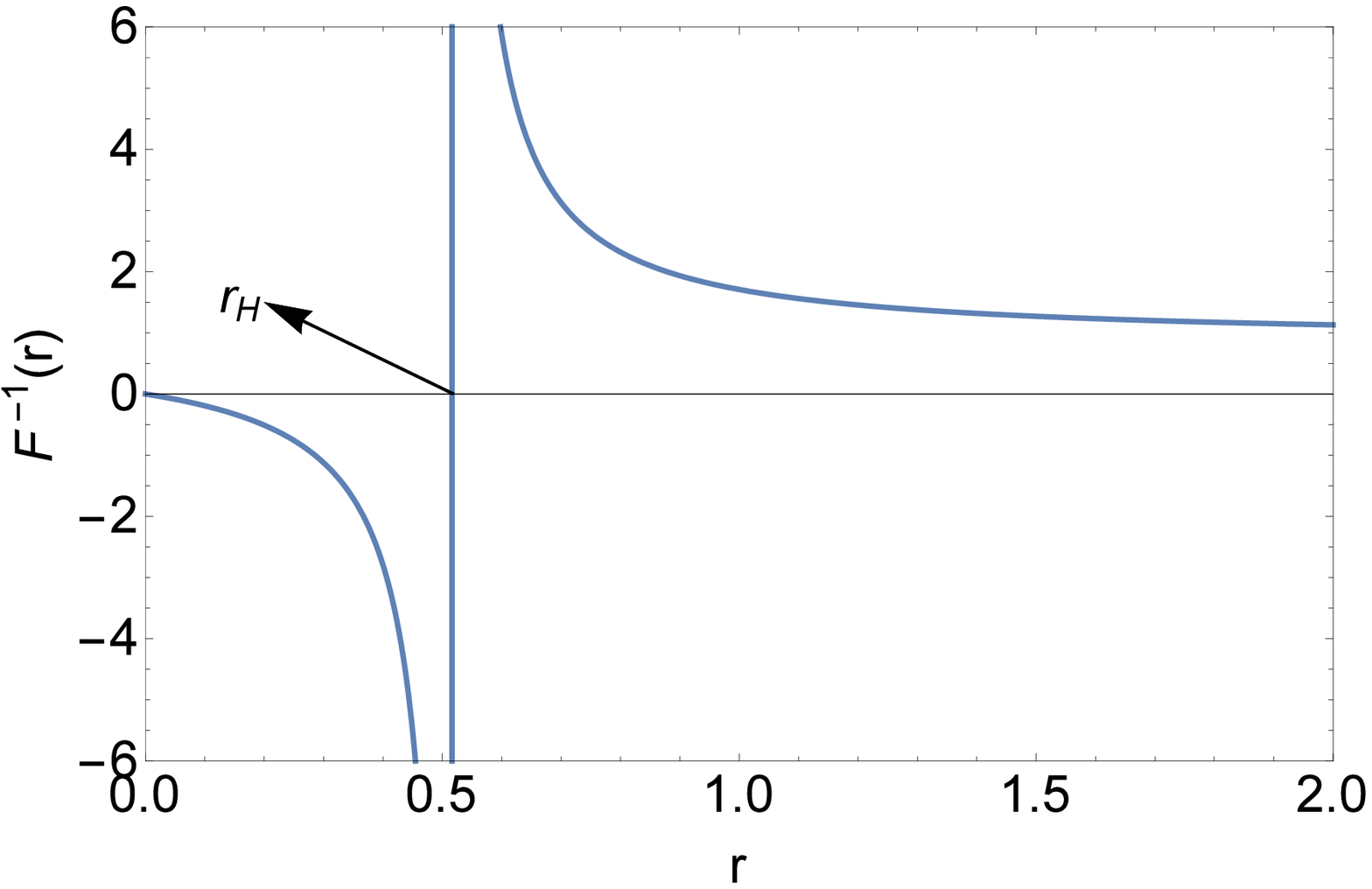}
\includegraphics[scale=0.41]{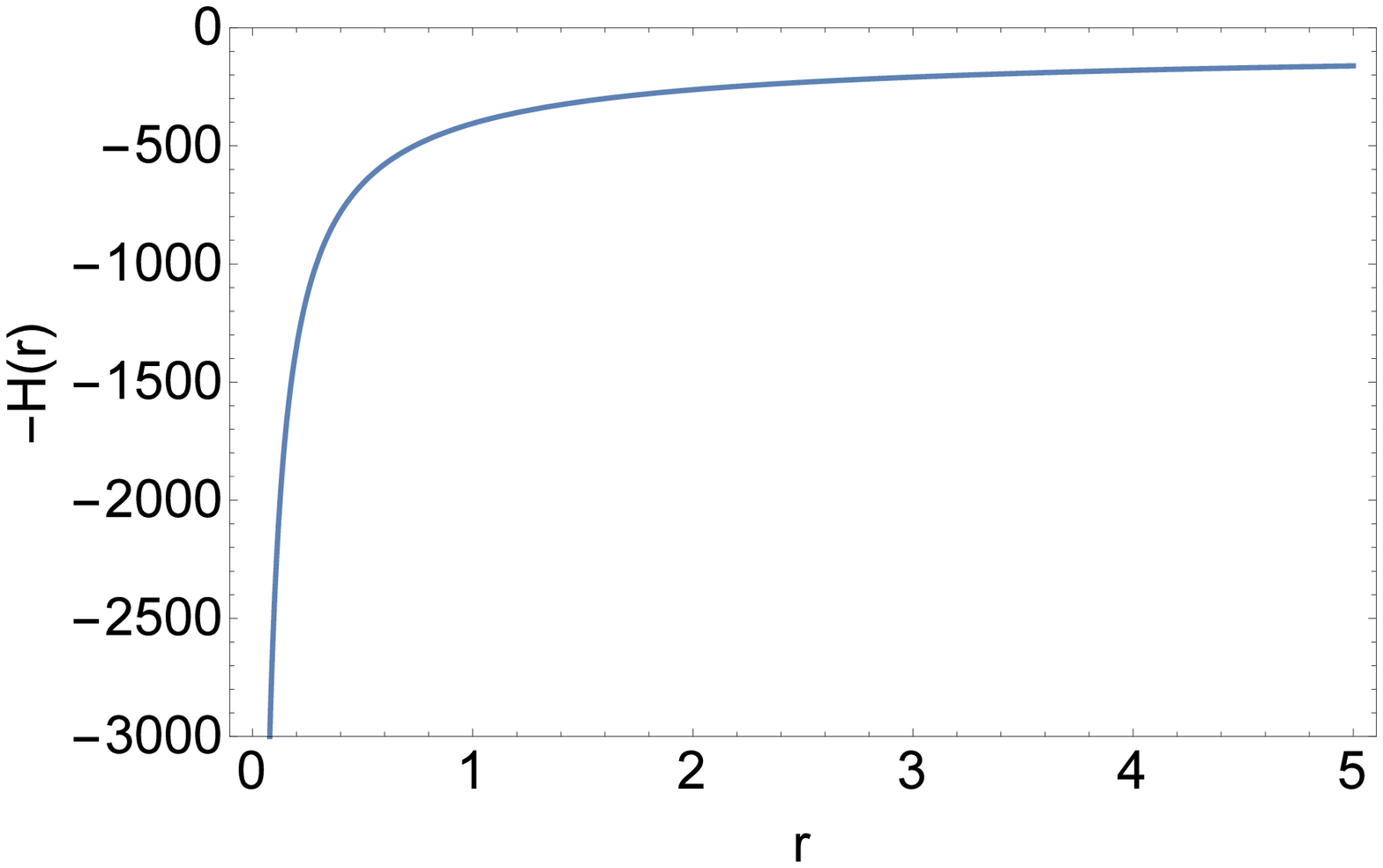}
\caption {Left panel: The behavior of radial component of spacetime metric for $a_1=10$, $c_1=1$, $c_2=-1$, $c_3=1$ and $b_1=-1/2$. The location of horizon is denoted by $r_{\sf H}$ . Right panel: The behavior of temporal component of spacetime metric for the same values as those of the left panel.}\label{firstsolh}
\end{figure}

\begin{figure}[htbp]
\hspace*{-1cm}
\includegraphics[scale=0.45]{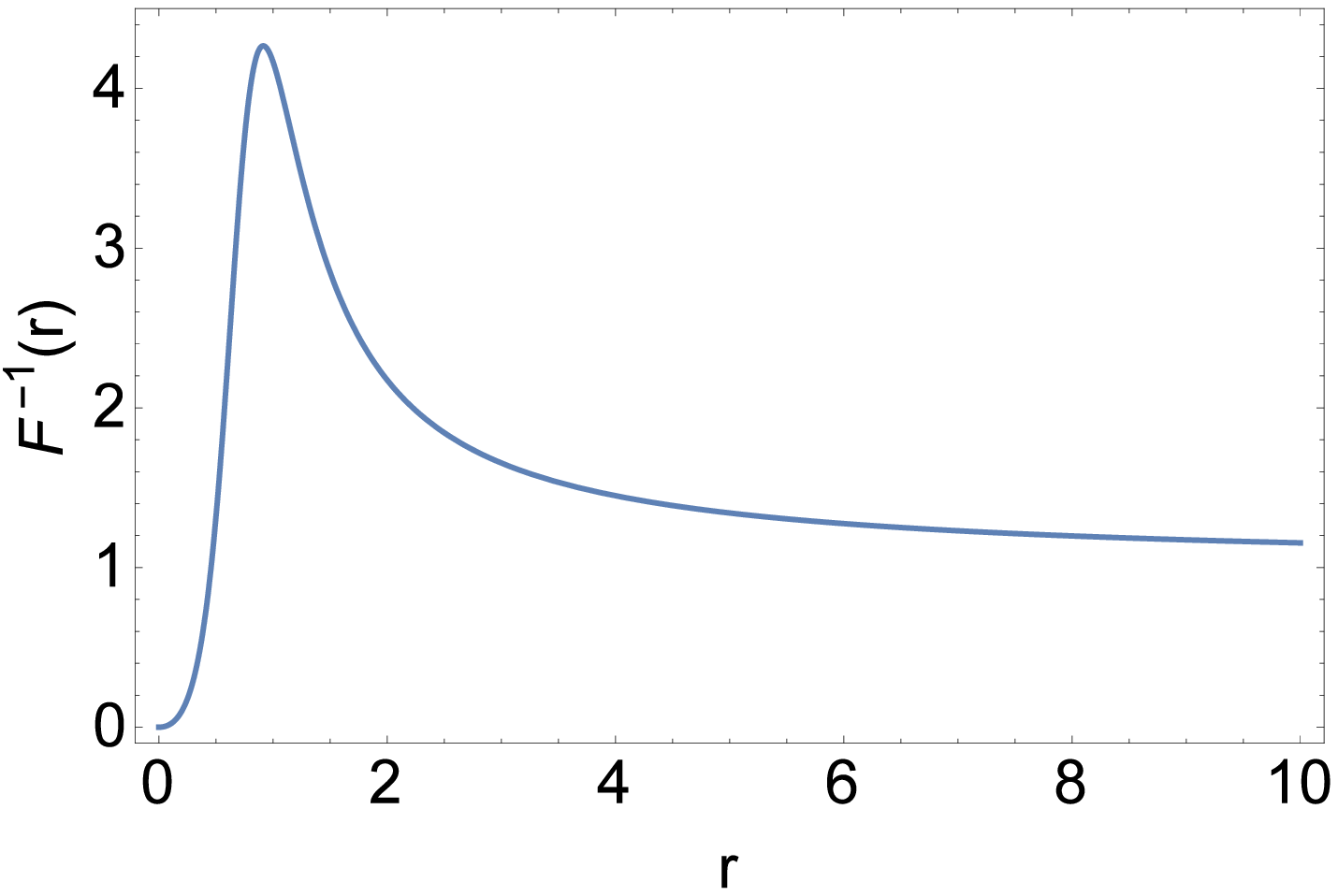}
\includegraphics[scale=0.47]{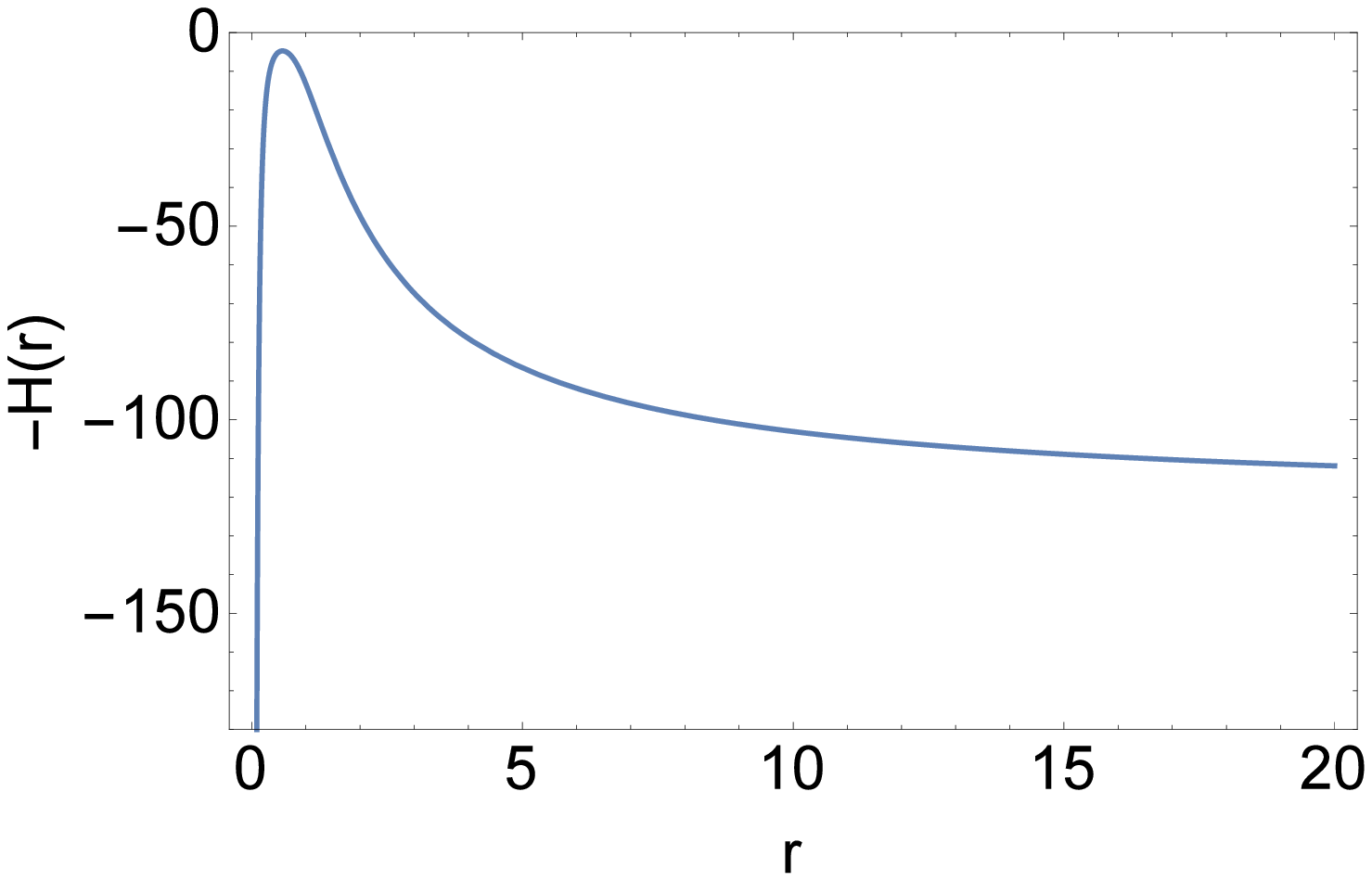}
\caption {Left panel: The behavior of radial component of spacetime metric for $m_0=0.7$ and $q_0=0.8$. Right panel: The behavior of temporal component of spacetime metric for the same values as those of the left panel and $a_1=10$ and $c_1=0.325$.}\label{second12}
\end{figure}

Let us begin with a static spherically symmetric line element given as
\be\label{lineelement}
ds^2=-H(r)dt^2+\f{dr^2}{F(r)}+r^2d\Omega^2,
\ee
where $d\Omega^2$ is the standard line element of a unit two-sphere. In general, it is possible to study spacetimes with 24 independent components of the torsion tensor. However, a physical interpretation of these components is very difficult. In order to find possible restrictions on the components of spacetime torsion, we consider simplifying assumptions using the symmetries of spacetime. For example, if $\xi^{\alpha}$ being a Killing vector field of the metric (\ref{lineelement}), i.e., $\mathscr{L}_\xi{\sf g}_{\mu\nu}=0$, we require that $\xi^{\alpha}$ leaves the torsion tensor invariant, i.e., $\mathscr{L}_\xi{\sf Q}^{\mu}_{\alpha\beta}=0$.  Such a restriction on the components of torsion has been proposed earlier in cosmological setting see e.g., \cite{restorcomp}. The Lie derivative of torsion leaves us with the following equation
\bea\label{Lietor}
&&\xi^{\rho}{\sf Q}^{\mu}_{\,\,\,\alpha\beta,\rho}-\xi^\mu_{\,\,,\rho} {\sf Q}^{\rho}_{\,\,\,\alpha\beta}+\xi^\rho_{\,\,,\alpha} {\sf Q}^{\mu}_{\,\,\rho\beta}+\xi^\rho_{\,\,,\beta} {\sf Q}^{\mu}_{\,\,\alpha\rho}+\xi^\lambda {\sf Q}^{\rho}_{\,\,\alpha\beta}{\sf Q}^{\mu}_{\lambda\rho}\nn&&+\xi^\lambda {\sf Q}^{\mu}_{\,\,\rho\beta}{\sf Q}^{\rho}_{\,\,\alpha\lambda}+\xi^\rho {\sf Q}^{\mu}_{\,\,\alpha\lambda}{\sf Q}^{\lambda}_{\,\,\beta\rho}=0.
\eea
The time-like and space-like Killing vector fields of metric (\ref{lineelement}) are given as, $\xi_{t}^\alpha=[1,0,0,0]$ and $\xi_{s}^\alpha=[0,0,0,1]$, respectively. It is then easy to show that if we take the non-vanishing components of the torsion tensor to be, ${\sf Q}^{\tt r}_{\,\,{\tt tr}}=A(r)$ and ${\sf Q}^\theta_{\,\,{\tt t}\theta}={\sf Q}^\phi_{\,\,{\tt t}\phi}=B(r)$, equation (\ref{Lietor}) will be satisfied for both Killing vector fields. Such a choice for torsion components has been also justified in \cite{PB1981}. The field equations (\ref{eqmtor-1}) and (\ref{eqmet-1}) in vacuum then read
\bea\label{fe00}
&&4a_1r^2HFH^{\prime\prime}(A^2+2B^2)-4a_1r^2FH^2(AA^{\prime\prime}+2BB^{\prime\prime})\nn
&-&5a_1r^2F(H^{\prime})^2(A^2+2B^2)+2a_1rHH^{\prime}\bigg[rF^\prime(A^2+2B^2)\nn&+&4F(rAA^\prime+2rBB^\prime+A^2+2B^2)\bigg]-12H\Bigg[\f{rHF^\prime}{6}\bigg(a_1r(AA^\prime+2BB^\prime)\nn&-&a_1(A^2+2B^2)-\f{H}{2}\bigg)+\f{a_1}{3}r^2FH\left((A^\prime)^2+2(B^\prime)^2\right)+\f{2a_1}{3}rFH\left(AA^\prime+2BB^\prime\right)\nn&+&\f{H^2}{12}(1-F)+H\Bigg(\f{B^2}{12}\left(4a_1-4a_1F-r^2\right)-\f{r^2}{6}AB+\f{a_1}{6}A^2(1-F)\Bigg)\nn&+&a_1r^2B\left(A^2+2B^2\right)\left(A+\f{B}{2}\right)\Bigg]=0,
\eea
as the $\texttt{[t,t]}$ component of (\ref{eqmet-1}),
\bea\label{trttor-01metricfh}
2a_1F(A+2B)\left[H^{\prime}\left(A^2+2B^2\right)-2H\left(AA^\prime+2BB^\prime\right)\right]=0,
\eea
as the $\texttt{[r,t]}$ component of (\ref{eqmet-1}),
\bea\label{ferr}
&&a_1r^2F(H^\prime)^2(A^2+2B^2)-2rFHH^\prime\Bigg[a_1r(AA^\prime+2BB^\prime)-a_1(A^2+2B^2)-\f{H}{2}\Bigg]\nn&+&H\left[2a_1(A^2+2B^2)-H\right]\big(H(1-F)+(2A+B)r^2B\big)\nn&-&8a_1rH^2F(AA^\prime+2BB^\prime)=0,
\eea
as the $\texttt{[r,r]}$ component of (\ref{eqmet-1}) and
\bea\label{festhethe}
&&2rFHH^{\prime\prime}\left[H+2a_1(A^2+2B^2)\right]-16a_1rFH^2(AA^{\prime\prime}+2BB^{\prime\prime})\nn&-&rF(H^\prime)^2\left[H+10a_1(A^2+2B^2)\right]+HH^\prime\Bigg\{rF^\prime\left(H+2a_1(A^2+2B^2)\right)\nn&+&F\Big[24a_1r(AA^\prime+2BB^\prime)+4a_1\left(A^2+2B^2\right)+2H\Big]\Bigg\}\nn&+&16H\Bigg\{-\f{H}{2}F^\prime\Bigg[a_1r(AA^\prime+2BB^\prime)+\f{a_1}{2}\left(A^2+2B^2\right)-\f{H}{4}\Bigg]\nn&-&ra_1FH\left[(A^\prime)^2+2(B^\prime)^2\right]-a_1FH\left[AA^\prime+2BB^\prime\right]\nn&+&rB\left(A+\f{B}{2}\right)\left(a_1[A^2+2B^2]-\f{H}{2}\right)\Bigg\}=0,
\eea
as the ${[\theta,\theta]}$ and ${[\phi,\phi]}$ components of (\ref{eqmet-1}). The field equations in vacuum for (\ref{eqmtor-1}) are obtained as
\bea\label{fqrrt}
&&2a_1r^2FAHH^{\prime\prime}-a_1r^2AF(H^\prime)^2+a_1rAHH^\prime\left(rF^\prime+4F\right)+4a_1rAH^2F^\prime+\nn&+&H^2\left[4a_1A(F-1)+2r^2B\right]-4a_1r^2B\left[3A^2+AB+2B^2\right]=0,
\eea
as the $\left[\texttt{r,r,t}\right]$ component of (\ref{eqmtor-1})
\bea\label{fqththt}
&2&\!\!\!\!\!\!a_1r^2FBHH^{\prime\prime}-a_1r^2FB(H^\prime)^2+a_1rBHH^\prime(rF^\prime+4F)-2H\Bigg\{-2a_1rBHF^\prime\nn&+&\!\!\!\!\!\!H\!\left[B\left(-\f{r^2}{2}-2a_1F\!+\!2a_1\right)-\f{r^2}{2}A\right]\!+\!a_1r^2\!\!\left(A^3\!+BA^2\!+6AB^2\!+4B^3\right)\!\!\!\Bigg\}=0,\nn
\eea
as the $\left[\theta,\theta,\texttt{t}\right]$ and $\left[\phi,\phi,\texttt{t}\right]$ components of (\ref{eqmtor-1}). We note that the $\left[\theta,\theta,\texttt{r}\right]$, $\left[\phi,\phi,\texttt{r}\right]$ and $\left[\texttt{t,t,r}\right]$ components of (\ref{eqmtor-1}) are proportional to (\ref{trttor-01metricfh}) with factor$F(A+2B)$. Solving equation (\ref{trttor-01metricfh}) for $H(r)$ we get\footnote{Note that, equation (\ref{trttor-01metricfh}) can be satisfied for $A=-2B$, as well. However, as we have considered the relation $A=b_1B$ between torsion components, we have put aside this special case.}
\bea\label{sol1}
H(r)=c_1\left(A^2+2B^2\right),
\eea
where $c_1$ is a constant of integration. Substituting for $H(r)$ into equations (\ref{fe00}), (\ref{fqrrt}) and (\ref{fqththt}) along with setting $A(r)=b_1B(r)$ and $c_1=2a_1$, we arrive at a single differential equation as
\bea\label{finaldiffeq}
2r^2FB^{\prime\prime}+\left[r^2F^\prime+4rF\right]B^\prime+\Bigg[2(F+rF^{\prime}-1)-\f{(2b_1+1)}{a_1(b_1^2+2)}r^2\Bigg] B=0,
\eea
where $b_1$ is dimensionless constant and with the above considerations, equations (\ref{ferr}) and (\ref{festhethe}) are satisfied correspondingly. The differential equation (\ref{finaldiffeq}) admits a general solution for the metric function $F(r)$, given as
\bea\label{gensolF(r)}
&&F(r)={\sf exp}\!\!\left[-\int\f{2(B+2rB^\prime+r^2B^{\prime\prime})}{r(2B+rB^\prime)}dr\right]\Bigg\{c_3+\nn&&\int\f{{\sf exp}\!\!\left[2\int\f{B+2rB^\prime+r^2B^{\prime\prime}}{r(2B+rB^\prime)}dr\right]B\left(2+\f{(2b_1+1)}{a_1(b_1^2+2)}r^2\right)}{r(2B+rB^\prime)}dr\Bigg\},
\eea
where $c_3$ is the integration constant. In order to find the metric functions $g_{\tt tt}(r)=-H(r)$ and $g_{\tt rr}(r)=1/F(r)$ we need to determine the functionality of $B(r)$. One way to simplify the integral (\ref{gensolF(r)}) is to consider the following differential equation for $B(r)$ as
\bea\label{diffeqforB}
B+2rB^\prime+r^2B^{\prime\prime}=n(2B+rB^\prime),
\eea
for which the solution reads
\bea\label{solbc1c2}
B(r)=r^{\f{1}{2}\left(n-1-s\right)}\left[c_1+c_2r^s\right],~~~~~~~s=(n(6+n)-3)^{\f{1}{2}},
\eea
where $c_1$ and $c_2$ are integration constants. Substituting for $B(r)$ back into the integral (\ref{gensolF(r)}) we find
\bea\label{finalsolFr}
F(r)&=&\f{c_3}{r^{2n}}+\f{1}{p}\Bigg[(n-s+3)\left(2a_1(n+1)(b_1^2+2)+n(1+2b_1)r^2\right)\nn&+&4sa_1(b_1^2+2)(1+n){}_2F_1\left[1,\f{2n}{s},1+\f{2n}{s},\f{c_2r^s(n+s+3)}{c_1(s-n-3)}\right]\nn&+&2ns(1+2b_1)r^2{}_2F_1\left[1,\f{2(n+1)}{s},\f{2+2n+s}{s},\f{c_2r^s(n+s+3)}{c_1(s-n-3)}\right]\Bigg],\nn
\eea
where ${}_2F_1$ is the hyper-geometric function and
\bea\label{consp}
p=a_1 \left(b_1^2+2\right) n (n+1) (n-s+3) (n+s+3).
\eea
Let us choose $n=\f{1}{2}$ for which we get
\bea\label{solforbr}
B(r)=c_2-\f{2c_1}{\sqrt{r}},
\eea
and
\bea\label{Fhalfint}
\!\!\!\!F(r)\!\!\!\!&=&\!\!\!\!\f{32a_1(b_1^2+2)c_2^4-27(2b_1+1)c_1^4}{32a_1(b_1^2+2)c_2^4}-\f{32a_1c_2^4(b_1^2+2)+81c_1^5(2b_1+1)}{32a_1(b_1^2+2)c_2^5\sqrt{r}}\nn&-&\!\!\!\!\f{3(2b_1+1)c_1^3\sqrt{r}}{8a_1c_2^3(b_1^2+2)}-\f{3c_1^2(2b_1+1)r}{16a_1(b_1^2+2)c_2^2}-\f{c_1(2b_1+1)r^{\f{3}{2}}}{10a_1c_2(b_2^2+2)}+\f{(2b_1+1)r^2}{6a_1(b_1^2+2)}\nn&+&\!\!\!\!\f{\left[64a_1c_3c_2^6(b_1^2+2)-3(81c_1^4(2b_1+1)+32a_1c_2^4(b_1^2+2))\ln(3c_1-2c_2\sqrt{r})\right]}{64a_1c_2^6(b_1^2+2)r},\nn
\eea
where we have redefined the constants $c_1$ and $c_2$. An asymptotically flat solution can be obtained by setting $b_1=-1/2$ as
\bea\label{frasymptot}
F(r)&=&1-\f{c_1}{c_2\sqrt{r}}+\f{6c_3c_2^2-9c_1^2\ln(3c_1-2c_2\sqrt{r})}{6c_2^2r},\\
H(r)&=&2a_1(b_1^2+2)\left[c_2-\f{2c_1}{\sqrt{r}}\right]^2.
\eea
The behavior of metric functions has been plotted in Fig. (\ref{firstsolh}). On the left panel we see the behavior of radial component of metric, the location of event horizon ($r_{\sf H}$) is determined by the condition $F(r)=0$. However, for $c_2<0$ and $c_1>0$, there is no infinite redshift surface as the temporal component of the metric never vanishes. For $r\rightarrow\infty$, the geometry becomes a Minkowski flat spacetime. For $r>r_{\sf H}$ the metric signature is Lorentzian, however, as we pass through the horizon ($r<r_{\sf H}$) the radial component becomes negative and the metric signature changes to $(--++)$, exhibiting a spacetime with 2+2-signature \cite{semieu,semieu1,semieu2,semieu3,semieu4,semieu5,semieu6}. Such a metric signature is physically interesting as it has applications in self-dual super-gravity and super-string theories \cite{SDSGSS,SDSGSS1,SDSGSS2,SDSGSS3,SDSGSS4,SDSGSS5,SDSGSS6,SDSGSS7,SDSGSS8,SDSGSS9}, the analysis of spinors in $2+2$ dimensions \cite{spinor2+2,spinor2+21}, cosmological models \cite{2+2cosmo} and black hole like solutions \cite{2+2BH}.
Another type of solutions can be found by setting
\bea\label{secondtype}
A(r)=b_1B(r),~~~H(r)=2a_1\left[A^2(r)+2B^2(r)\right],~~~F(r)=1-\f{2m_0}{r}+\f{q_0^2}{r^2},
\eea
where $q_0$ and $m_0$ are constants. We then get the following differential equation for $b_1=-1/2$,  as
\bea\label{diffeqB} 
q_0^2B-rB^{\prime}(q_0^2-3m_0r+2r^2)-r^2B^{\prime\prime}(q_0^2-2m_0r+r^2)=0,
\eea
for which the solution reads
\bea\label{solBCOSH}
B(r)=c_4\frac{\Bigg\{\left(m_0^2+1\right) r^2+q_0^2 r (r-4 m_0)-q_0 r \xi(r)^{\f{1}{2}}[2q_0^2-2m_0r]+2 q_0^4\Bigg\}}{2 r \left[q_0 \left(\xi(r)^{\f{1}{2}}+q_0\right)-m_0 r\right]},
\eea
where $c_4$ is an integration constant and $\xi(r)=r (r-2 m_0)+q_0^2$. From the above expression and the second part of (\ref{secondtype}) we can find the temporal component of metric as $H(r)=(9/2)a_1B^2(r)$. This solution is asymptotically flat for a suitable choice of $c_4$ as given by
\bea\label{assflatHr}
H(r)\Big|_{r\rightarrow\infty}\!\!\!\!\!\!=1+\f{\eta}{\zeta r}+\f{\Upsilon}{\zeta^2r^2}+{\mathcal O}\left(\f{1}{r^3}\right),
\eea
where
\bea\label{coeffeszeta}
\eta\!\!\!\!&=&\!\!\!2q_0(m_0^2+q_0^2-2m_0q_0-1),~~~\zeta=m_0^2+q_0^2-2m_0q_0+1,\nn
\Upsilon\!\!\!\!&=&\!\!\!q_0^5(2q_0-7m_0)+2q_0^3m_0^2(4q_0-m_0)+q_0(m_0^4-1)(m_0-2q_0),
\eea
and we have set
\bea\label{c4assfl}
c_4=\pm\f{2\sqrt{2}}{3}\left\{\f{a_1[(m_0-q_0)^2+1]^2}{(m_0-q_0)^2}\right\}^{-\f{1}{2}}.
\eea
From the third part of (\ref{secondtype}) the location of the horizons is given by the condition $F(r_{{\sf H}})=0$. We then observe that the sapcetime could admit two horizons locating at 
\bea\label{hors54}
r^{\pm}_{{\sf H}}=m_0\pm\sqrt{m_0^2-q_0^2}.
\eea
For $m_0>q_0$ the radius of horizons is real, however, for these values of $r^{\pm}_{{\sf H}}$, equation $\xi(r_{{\sf H}})=0$ can be rewritten as
\bea\label{rHxi}
\left[r^{+}_{{\sf H}}-(m_0+\sqrt{m_0^2-q_0^2})\right]\left[r^{-}_{\sf H}-(m_0-\sqrt{m_0^2-q_0^2})\right]=0,
\eea
from which we see that $\xi(r)$ function will have two real roots. This means that $\xi(r)$ is negative between the roots as the coefficient of $r^2$ is positive. A negative value for $\xi(r)$ leads to a non-zero imaginary part for $B(r)$ and therefore makes the $H(r)$ function to get complex values which is unphysical. In case in which, $m_0\rightarrow q_0$, the constant $c_4$ tends to infinity and this contradicts the asymptotic flatness. We are therefore left with the only case, i.e., $m_0<q_0$ for which the spacetime is free of horizon and $\xi(r)$ function will no longer have any real root. Therefore it stays posotive for all values of $r$ and the $H(r)$ function will take real values throughout the spacetime. Figure (\ref{second12}) shows the behavior of metric components as a function of radial coordinate. It is seen that the radial component is always positive and the temporal one is negative so that the signature remains Lorentzian throughout the spacetime.

\begin{figure}[htbp]
\hspace*{-1cm}
\includegraphics[scale=0.429]{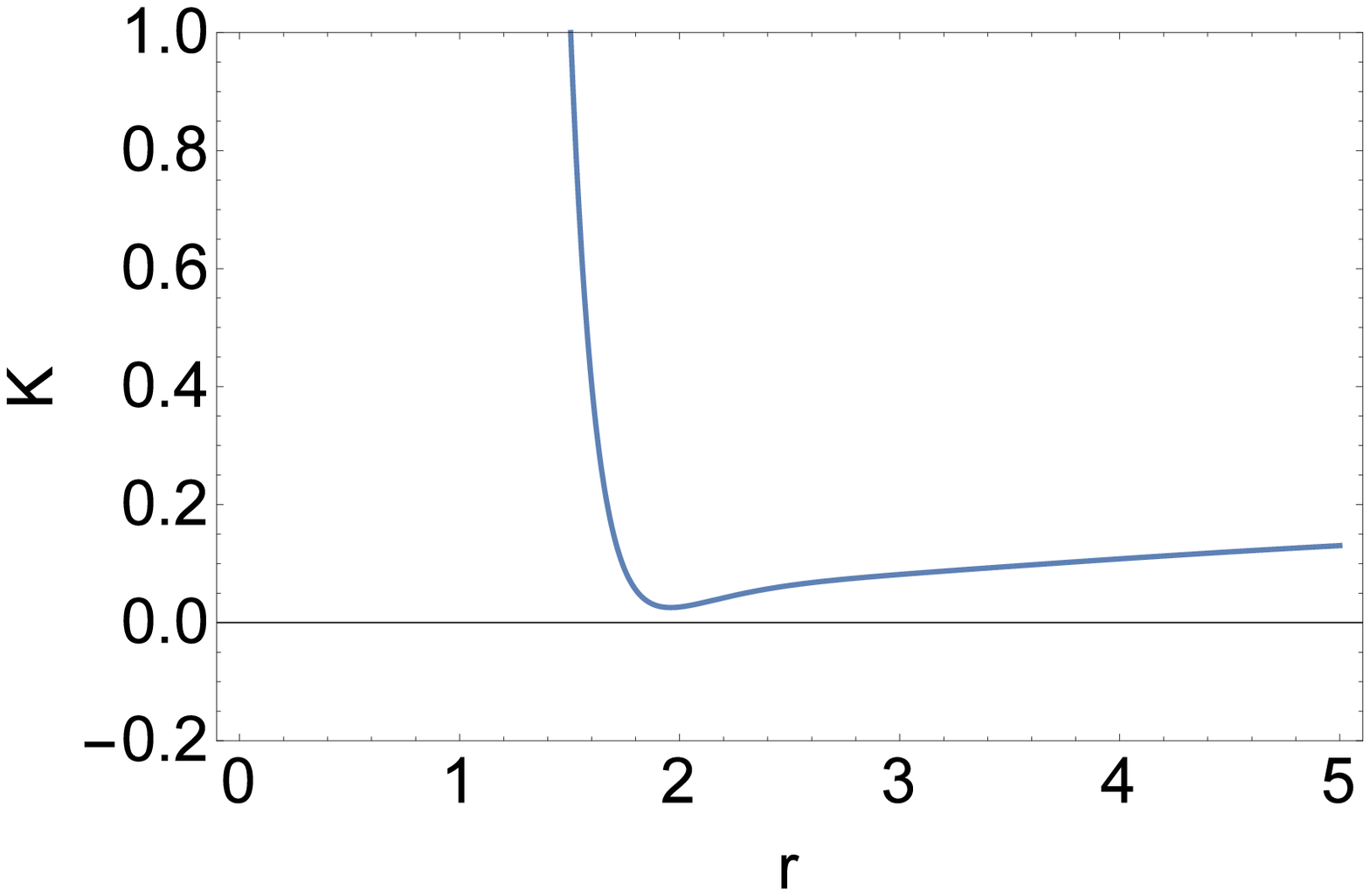}
\includegraphics[scale=0.418]{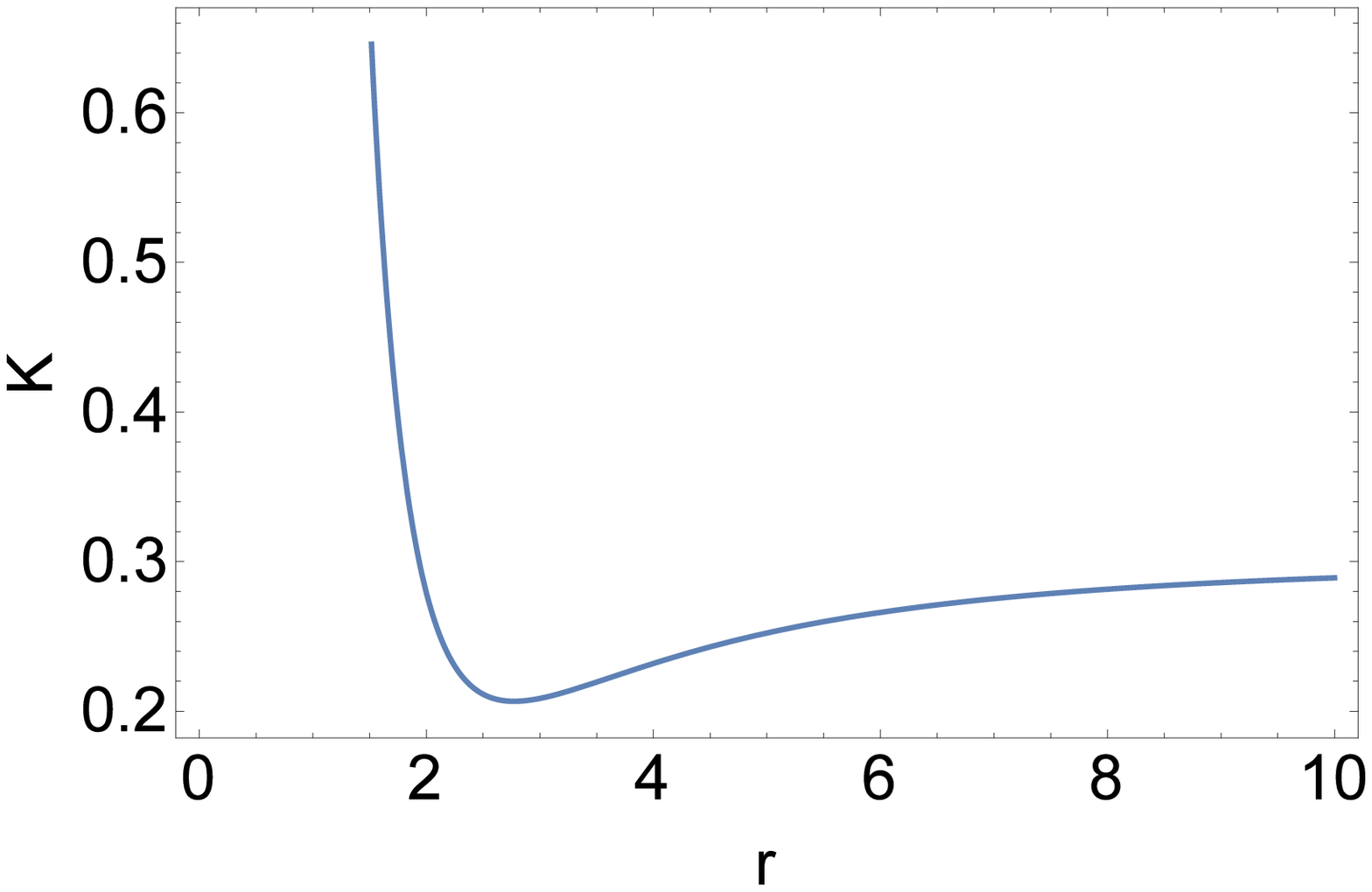}
\caption {Left panel: The behavior of Kretschmann scalar for $a_1=10$, $c_1=1$, $c_2=-1$, $c_3=1$ and $b_1=-1/2$. Right panel: The behavior of Kretschmann scalar for $m_0=0.7$ and $q_0=0.8$, $a_1=10$ and $c_1=0.325$.}\label{KRSC12}
\end{figure}

In figure (\ref{KRSC12}) we have plotted the Kretschmann invariant\footnote{The Riemann curvature tensor of affine connection can be found as 
\bea\label{RiCurTen}
\tilde{{\sf R}}^{\lambda}_{\,\,\,\mu\nu\rho}={\sf R}^{\lambda}_{\,\,\,\mu\nu\rho}+\nabla_{\nu}{\sf K}^{\lambda}_{\,\,\mu\rho}-\nabla_{\rho}{\sf K}^{\lambda}_{\,\,\mu\nu}+{\sf K}^{\sigma}_{\,\,\mu\rho}{\sf K}^{\lambda}_{\,\,\sigma\nu}-{\sf K}^{\sigma}_{\,\,\mu\nu}{\sf K}^{\lambda}_{\,\,\sigma\rho},
\eea
whence we can find the Kretschmann scalar through the following relation
\bea\label{KRET}
\tilde{{\sf K}}=\tilde{{\sf R}}^{\alpha\beta\gamma\delta}\tilde{{\sf R}}_{\alpha\beta\gamma\delta}.
\eea} for the two solutions presented in this subsection. In the left panel, this quantity is finite except in the limit of approach to $r=0$ where it diverges. This means that we have a curvature singularity when $r$ vanishes. However, this singularity is necessarily covered by a spacetime event horizon. In the right panel, the Kretschmann scalar is plotted for the second case where we see again this quantity is regular throughout the spacetime and diverges in the limit $r\rightarrow0$, signaling the existence of a curvature singularity at this point. However, the singularity is naked in contrast to a black hole, where it is hidden behind an event horizon. In such a situation, there can be light rays terminating at the singularity reaching faraway observers in the universe exposing thus, the ultra strong gravity regimes to such observers \cite{Joshibook}. One way to pursue this issue is to investigate the behavior of timelike and null geodesics around the singularity which could provide useful astrophysical information about the nature of singularity and the interactions with its surrounding medium \cite{nakedinfomedium,nakedinfomedium1,nakedinfomedium2,nakedinfomedium3,nakedinfomedium4,nakedinfomedium5,nakedinfomedium6,nakedinfomedium7,nakedinfomedium8}.
\subsection{Class B Solutions}\label{IIB}
For this class of solutions we parametrize the line element as
\bea\label{ClassBmetric}
ds^2=-H(r)dt^2+\f{dr^2}{H(r)}+r^2d\Omega^2.
\eea
The field equations (\ref{eqmtor-1}) and (\ref{eqmet-1}) in vacuum then read
\bea\label{00metric}
&&H\Bigg[H\left(1-rH'\right)-H^{2}-r^{2}\left(B^{2}+2AB\right)\Bigg]+a_{1}\Bigg\{r^2\bigg[-4HH^{\prime\prime} \left(A^2+2B^2\right) \nn
&+&4H^2AA^{\prime\prime}+8H^2BB^{\prime\prime}+3\left(H^\prime\right)^2 \left(A^2+2B^2\right)^{2}\bigg]\nn
\nn
&-&6\Bigg[ r\bigg(AA^\prime+2BB^\prime\bigg)+\f{5}{3}\bigg(A^2+2B^{2}\bigg)\Bigg]r H H^\prime+4r^2H^{2}\left(A^\prime\right)^{2}\nn
&+&8rH^{2}AA^\prime+8r^2H^{2}\left(B^\prime \right)^{2}+16rH^2BB^\prime 
-2(A^2+2B^{2})H^{2}\nn&+&\hspace{-.2cm} 2\left(A^2+2B^2\right)H +12r^2B\left(A+\f{1}{2}B\right)\left(A^2+2B^2\right) \Bigg\}=0,\nn
\eea

as the $\texttt{[t,t]}$ component of (\ref{eqmet-1}),
\bea\label{trttor-01metric}
a_{1}(A+2B)\Bigg[H^{\prime}\left(A^2+2B^2\right)-2H\left(AA^\prime+2BB^\prime\right)\Bigg]=0,
\eea
as the $\texttt{[r,t]}$ component of (\ref{eqmet-1}),
\bea\label{11metric}
&-&2H\left[Br^2\left(A+\f{B}{2}\right)+\f{H}{2}(1-H)\right]+rH^{2}H'\nn
&+&a_1\Bigg\{4\left(A^2+2B^2\right)\left[Br^2\left(A+\f{B}{2}\right)+\f{H}{2}(1-H)\right]\nn
&-&2rHH^\prime\left[r\left(AA^\prime+2BB^\prime\right)-\left(A^2+2B^2\right)\right]\nn
&+&r^2\left(H^\prime\right)^2\left(A^2+2B^2\right)-8rH^2\left(AA^\prime+2BB^\prime\right)\Bigg\}=0,\nn
\eea
as the $\texttt{[r,r]}$ component of (\ref{eqmet-1}) and
\bea\label{22metric}
&&rH^{2}H^{\prime\prime}+2H^{2}H'-4rHB\left(A+\f{B}{2}\right)+a_{1}\Bigg\{2rHH^{\prime\prime}\left(A^2+2B^2\right)\nn
&-&8rH^2\left(AA^{\prime\prime}+2BB^{\prime\prime}\right)-4r\left(H'\right)^2\left(A^2+2B^2\right)+8rHH'\left(AA'+2BB'\right)\nn
&-&8rH^2\left[(A^\prime)^2+2(B^\prime)^2\right]-8H^2\left[AA^\prime+2BB^\prime\right]\nn&+&8rB\left(A^2+2B^2\right)\left(A+\f{B}{2}\right)\Bigg\}=0,
\eea
as the ${[\theta,\theta]}$ and ${[\phi,\phi]}$ components of (\ref{eqmet-1}). The field equations for (\ref{eqmtor-1}) are obtained as

\bea\label{rrttor}
&-&Br^{2}H+a_{1}\Bigg[-r^2AHH^{\prime\prime}-4rAHH^\prime+6r^2BA^2\nn
&+&2A\Big(r^2B^2-H^2+H\Big)+4B^3r^2\Bigg]=0,
\eea
as the $\left[\texttt{r,r,t}\right]$ component of (\ref{eqmtor-1})
\bea\label{ththttor}
&-&Br^2H-r^2AH+a_{1}\Bigg[-2r^2BHH^{\prime\prime}-8rBHH^\prime+8r^2B^3\nn
&+&12r^2AB^2+B\left(2r^2A^2-4H^2+4H\right)+2r^2A^3\Bigg]=0,
\eea
as the $\left[\theta,\theta,\texttt{t}\right]$ and $\left[\phi,\phi,\texttt{t}\right]$ components of (\ref{eqmtor-1}). The $\left[\theta,\theta,\texttt{r}\right]$, $\left[\phi,\phi,\texttt{r}\right]$ and $\left[\texttt{t,r,t}\right]$ components of (\ref{eqmtor-1}) are the same as (\ref{trttor-01metric}). Solving equation (\ref{trttor-01metric}) for $H(r)$ we get
\bea\label{sol1}
H(r)=c_1\left[A(r)^2+2B(r)^2\right],
\eea
where $c_1$ is a constant of integration. Substituting $H(r)$ into equations (\ref{00metric}), (\ref{11metric}) and (\ref{22metric})-(\ref{ththttor}) together with setting $A(r)=b_0B(r)$ and $c_1=2a_1$, we arrive at a single differential equation as
\bea\label{singleeq}
&&4a_1(b_1^2+2)\left[BB^{\prime\prime}+\left(B^{\prime}\right)^2\right]+\f{16a_1}{r}(b_0^2+2)BB^\prime+\f{4a_1}{r^2}(b_0^2+2)B^2\nn&-&\f{(2b_0+1)}{a_1(b_0^2+2)}-\f{2}{r^2}=0,
\eea
where $b_0$ is dimensionless constant. Solving the above differential equation, we get
\bea\label{solFC}
B(r)&=&\pm\f{1}{12a_1(b_0^2+2)r}\bigg[6r^4(2b_0+1)+72a_1(b_0^2+2)r^2\nn&-&288c_2a_1^2(b_0^2+2)^2r
+288c_3a_1^2(b_0^2+2)^2\bigg]^{\f{1}{2}},
\eea
where $c_2$ and $c_3$ are constants of integration. Substituting the above solution back into equation (\ref{sol1}) we finally get the metric function as
\bea\label{metrfunc}
H(r)=1-4\f{a_1c_2}{r}(b_0^2+2)+4\f{a_1c_3}{r^2}(b_0^2+2)+\f{b_0+\f{1}{2}}{6a_1(b_0^2+2)}r^2.
\eea
For $b_0=-1/2$, and $a_1$, $c_2$ and $c_3$ being positive\footnote{In case we set $c_3<0$, we have to take $a_1$ and $c_2$ to be negative too, in order to avoid negative mass and imaginary charge.}, the above solution represents a Reissner-Nordstrom black hole with mass and charge given as
\be\label{masscharge}
{\sf M}=(9a_1c_2/2{\sf G}){\sf c}^2,~~~~{\sf q}^2=(36\pi/{\sf G})a_1c_3\epsilon_0{\sf c}^4,
\ee
respectively. We therefore see that even in vacuum, the interaction between torsion and curvature could lead to a charged black hole solution. The location of event horizon can be obtained by setting $H(r)=0$ which gives
\be\label{horloc}
r^{\pm}_{{\sf H}}=\f{9}{2}a_1c_2\left[1\pm\left(1-\f{4}{9}\f{c_3}{a_1c_2^2}\right)^{\f{1}{2}}\right].
\ee
For $a_1>4c_3/9c_2^2$, the metric possesses two coordinate singularities at $r^+_{{\sf H}}$ and $r^-_{{\sf H}}$ which indeed exhibit the location of two horizons. In case where $a_1<4c_3/9c_2^2$ there are no event horizons to prevent faraway observers from detecting the curvature singularity at $r=0$, signaling that the singularity is naked. Hence, the spacetime torsion could affect the causal structure of spacetime and radius of the horizons. On the other hand, if the singularity is naked, the regimes of extreme gravity would be in causal connection with the observers in the universe and thus proving a suitable setting to detect the footprint of spacetime torsion. For $c_3=0$, and $b_0\neq-1/2$, solution (\ref{metrfunc}) represents the exterior field of a spherical mass (or black hole) in a de-Sitter or anti-de Sitter background, depending on the sign of $a_1$ and $b_0$ so that the cosmological constant is given by
\be\label{lambdavacuum}
\Lambda=(2b_0+1)/\left[4a_1(b_0^2+2)\right].
\ee
In case in which $0<9{\sf G}^2{\sf M}^2{\sf c}^2\Lambda<1$, the horizon equation
\bea\label{horeqschde}
\frac{6 {\sf G} {\sf M}}{{\sf c}^2 \Lambda }-\frac{3 r}{\Lambda }+r^3=0
\eea
admits two distinct positive real roots at $r = r_1$ and $r = r_2$ so that $r_2 > r_1 > 0$ with the third root being negative.
The smaller root, $r_1$ corresponds to a black hole-type event horizon, while the
larger root $r_2$ corresponds to a cosmological de Sitter-type event horizon \cite{gibhawdesitter}. In this case we observe that the spacetime torsion could play the role of cosmological constant through its coupling with curvature. For $9{\sf G}^2{\sf M}^2{\sf c}^2\Lambda=1$, the black hole and cosmological event horizons coincide at a null hyper-surface with radius $1/\sqrt{\Lambda}$. This class of solution is known as the Nariai solution \cite{Nariaisol}. Finally for $9{\sf G}^2{\sf M}^2{\sf c}^2\Lambda>1$, there is no real positive root for equation (\ref{horeqschde}) and thus the curvature
singularity at $r = 0$ is actually naked.  We note that once we set $a_1$ to be zero, the action (\ref{ECmod1}) reduces to the standard {\sf ECKS} action. We can therefore observe that when $a_1$ is zero, the field equation (\ref{eqmtor-1}) in vacuum implies a vanishing torsion field or correspondingly, $A(r)=B(r)=0$. We then get for the rest of field equations the following differential equation for metric function as
\be\label{coupconszero}
rH^{\prime\prime}+2H^{\prime}=0,
\ee
for which we readily find the solution  
\be\label{cczschw}
H(r)=c_4+\f{c_5}{r}.
\ee
This is nothing but the well-known Schwarzschild solution, once we set $c_4=1$ and $c_5=2{\sf GM}/{\sf c}^2$.
\section{Concluding remarks}\label{conclu}
In the present work we tried to obtain static vacuum spacetimes in generalized {\sf ECKS} theory by considering higher order terms from curvature and torsion within the gravitational Lagrangian. As we know in {\sf ECKS} theory, gravitational interactions are not only mediated by spacetime curvature but also such interactions could be due to the spacetime torsion. However these two fields could only interact through the spacetime metric which makes the spacetime torsion to be a non-dynamical field in {\sf ECKS} gravity as it obeys a pure algebraic field equation. It can therefore be of interest to consider Lagrangians that allow for non-minimal coupling between spacetime torsion and curvature, like scalar tensor theories of gravity \cite{sttg}. As a result of direct coupling with curvature, the field equations contain derivatives of spacetime torsion and thus making the torsion tensor to propagate even in the absence of matter fields.  Therefore, the inclusion of interacting terms between curvature and torsion could provide a framework within which the spacetime curvature and torsion would mutually affect each other and thus, even in vacuum, leading to non-trivial spacetimes as solutions to the field equations. Such a setting may be helpful for seeking the possible effects of spacetime torsion within the gravitational interactions.

Two classes of solutions we have found show that both black holes and naked singularities could emerge in vacuum. The first one deals with static spherically symmetric spacetimes admitting curvature singularity that could be either hidden behind the event horizon or can be seen by observers in the universe. The second class of solutions correspond to the well-known static solutions in {\sf GR} in the presence of e.g. electric field. These solutions also admit spacetime singularities either cloaked within an event horizon of gravity or visible to the external universe. From the observational perspective, a black hole can be distinguished from a naked singularity owing to the physical properties of their accretion disks that may form around them \cite{obsnakedbh,obsnakedbh1,obsnakedbh2}. Moreover, gravitational
lensing effects as a result of extreme curvature regions provide astronomers with a suitable tool to search for the
observational signatures coming out from a naked singularity so that lensing
characteristics of such objects are qualitatively very different from those formed as
Schwarzschild black holes \cite{nakedvsblhole,nakedvsblhole1,nakedvsblhole2,nakedvsblhole3,nakedvsblhole4}. From this point of view, it would be interesting to consider the equation
of photon trajectories around these objects and study the different aspects of lensing effects of naked singularities and black holes in oder to possibly detect the effects of spacetime torsion. However, dealing with this issue is out of the scope of the present work.

Finally, as we near to close this paper it deserves to point out a few notes on torsion gravity theories. Not long ago, spacetime torsion, in the context of {\sf GR}, did not seem to provide cosmological models with observational signatures as the phenomena including intrinsic angular momentum of fermionic particles and gravitation were considered to be significant only in the realm of very early Universe~\cite{ecksbheui}-\cite{gasperiniavoid}. However, it has been proven that spin is not the only source for spacetime torsion and in fact torsion can be decomposed into three irreducible parts, with different properties. A beautiful discussion on this issue can be found in~\cite{Annalsspin} where a systematic classification of these different types of torsion and their possible sources is surveyed. Based on geometrical classification of torsion tensors as provided in~\cite{Annalsspin} a wide class of torsion models could be studied independently of spin as their source. Moreover, an alternative approach to gravitational interaction named teleparallel gravity has been presented in~\cite{teleparallel,teleparallel1,teleparallel2,teleparallel3,teleparallel4} which corresponds to a gauge theory for the
translation group. Recently, new dynamical degrees of freedom have been introduced inside the teleparallel scheme and as a result more general Lagrangians including nonlinear functions of torsion scalar have been introduced~\cite{fTLags,fTLags1,fTLags2,fTLags3,fTLags4}. Cosmological as well as astrophysical applications of these type of gravity theories have been studied in~\cite{fTgrav} where it is shown that beside spherically symmetric and black hole solutions, a theoretical interpretation of the late-time acceleration of the Universe can be constructed. Work along this line has been perused to seek for other couplings of spacetime torsion to e.g., scalar fields~\cite{extendfTgrav}. In this sense, the form of teleparallel Lagrangian has been extended to include a scalar field and its kinetic term within the Lagrangian. In comparison with above works, the study of torsion theories without resorting to a spin fluid source was the aim of the present paper and in the context of the present model, it was shown that nontrivial spacetimes as the exact solutions to the field equations could be obtained. However, the inclusion of a spin source could possibly provide a richer framework by the virtue of which one is able to search for the footprints of torsion within the gravitational phenomena and indeed, this signals the importance of including torsion within a gravitation theory.



\begin{thebibliography}{99}
\bibitem{grtest} C. M. Will, Living Rev. Rel. {\bf 17} 4 (2014).
\bibitem{gwave} Virgo and LIGO Scientific collaborations, B. P. Abbott et al., Phys. Rev. Lett. {\bf 116} 061102 (2016).
\bibitem{obsertest} C. M. Will, \lq\lq{}{\it Theory and Experiment in Gravitational Physics}\rq\rq{}, Cambridge University Press, (1993).
\bibitem{obsertest1} E. Poisson and C. M. Will, \lq\lq{}{\it Gravity: Newtonian, Post-Newtonian, Relativistic}\rq\rq{}, Cambridge University Press, (2014).
\bibitem{infdedm} S. Capozziello and M. De Laurentis, Phys. Rept. {\bf 509} 167 (2011).
\bibitem{infdedm1} A. De Felice and S. Tsujikawa, Living Rev. Rel. {\bf 13} 3 (2010).
\bibitem{infdedm2} E. Berti et al., Class. Quant. Grav. {\bf 32} 243001 (2015).
\bibitem{infdedm3} P. Avelino et al., Symmetry {\bf 8} 70 (2016).
\bibitem{strbhneugr} I. H. Stairs, Living Rev. Relativity, {\bf 6} 2003-5 (2003).
\bibitem{DaEsFa} T. Damour and G. Esposito-Farese, Phys. Rev. Lett., {\bf 70} 2220 (1993); Phys. Rev. D {\bf 54} 1474 (1996).
\bibitem{hehlvonkerlick} F. W. Hehl, P. von der Heyde, and G. D. Kerlick and J. M. Nester, Rev. Mod. Phys. \textbf{48} (1976) 393.
\bibitem{PAPSPIn} A. Papapetrou, Philosphical Magazine, {\bf 40} 937 (1949).
\bibitem{WEYRABB} J. Weyssenhoff and A. Raabe, Acta Phys. Pol., {\bf 9} 19 (1947).
\bibitem{ecartan} E. Cartan, C. R. Acad. Sci. (Paris) {\bf 174} 593 (1922).
\bibitem{ecartan1} E. Cartan, Ann. Ec. Norm. Sup. {\bf 40} 325 (1923).
\bibitem{ecartan2} E. Cartan, Ann. Ec. Norm. Sup. {\bf 41} 1 (1924).
\bibitem{ecartan3} E. Cartan, Ann. Ec. Norm. Sup. {\bf 42} 17 (1925).
\bibitem{KibScia} T. W. B. Kibble, J. Math. Phys. {\bf 2} 212 (1961).
\bibitem{KibScia1} D. W. Sciama, in Recent Developments in General Relativity (Pergamon+PWN, Oxford, UK, 1962) p. 415.
\bibitem{KibScia2} D. W. Sciama, Rev. Mod. Phys. {\bf 36} 463 (1964).
\bibitem{KibScia3} F. Hehl and E. Kroner, Z. Phys. {\bf 187} 478 (1965).
\bibitem{KibScia4} F. Hehl, Abb. Braunschweig. Wiss. Ges.{\bf 18} 98 (1966).
\bibitem{KibScia5} A. Trautman, Bull. Polon. Acad. Sci. {\bf 20} 185 (1972).
\bibitem{KibScia6} A. Trautman, Bull. Polon. Acad. Sci. {\bf 20} 503 (1972).
\bibitem{ecksbheui} A. Trautman, Nature Physical Science {\bf 242} 7 (1973);\\ J. Stewart and P. Hajicek, Nature Physical Science {\bf 244} 96 (1973).
\bibitem{Venzo-Hehl} V. De Sabbata and M. Gasperini, \lq\lq{}{\it Introduction to Gravitation},\rq\rq{} World Scientific (1985).
\bibitem{Venzo-Hehl1} V. de Sabbata  and C. Sivaram, \lq\lq{}{\it Spin and Torsion in Gravitation},\rq\rq{} World Scientific (1994).
\bibitem{Venzo-Hehl2} F. W. Hehl, P. von der Heyde and G. David Kerlick, Rev. Mod. Phys {\bf 48} 393 (1976).
\bibitem{Venzo-Hehl3} F. W. Hehl, Gen. Relativ. Grav. {\bf 4} 333 (1973); {\bf 5} 491 (1974).
\bibitem{Venzo-Hehl4} P. G. Bergmann and V. De Sabbata, (eds.) \lq\lq{}{\it Cosmology and Gravitation: Spin, Torsion, Rotation, and Supergravity},\rq\rq{} Springer Science \& Business Media (2012).
\bibitem{Gasper} M. Gasperini, Phys. Rev. Lett. {\bf 56} 2873 (1986).
\bibitem{gasperiniavoid} M. Gasperini, Gen. Rel. Grav. {\bf 30} 1703 (1998).
\bibitem{gasperiniavoid1}  N. J. Poplawski, Phys. Rev. D, {\bf 85} 107502 (2012).
\bibitem{gasperiniavoid10} N. J. Poplawski, Gen. Relativ. and Gravit., {\bf 44} 1007 (2012).
\bibitem{gasperiniavoid100} N. J. Poplawski,  Phys. Lett. B, {\bf 690} 73 (2010).
\bibitem{gasperiniavoid2} S. D. Brechet, M. P. Hobson, A. N. Lasenby, Class. Quantum Grav. {\bf 25} 245016 (2008).
\bibitem{torquantum} L. Freidel, D. Minic and T. Takeuchi, Phys. Rev. D {\bf 72} 104002 (2005).
\bibitem{torquantum1} C. Pagani and R. Percacci, Class. Quantum Grav. {\bf 32} 195019 (2015).
\bibitem{torquantum2} T. P. Singh, Current Science {\bf 109} 2258 (2015)
\bibitem{torquantum3} A. J. Hanson and T. Regge, Group Theoretical Methods in Physics, {\bf 94} 354 (1979);\\ I. L. Shapiro, Phys. Rep. {\bf 357} 113 (2002).
\bibitem{Putzfield} D. Puetzfeld, New Astronomy Reviews {\bf 49} 59 (2005).
\bibitem{emerun} H. Hadi, Y. Heydarzade, M. Hashemi and F. Darabi, Eur. Phys. J. C {\bf 78} 38 (2018).
\bibitem{grcolapse} M. Hashemi, S. Jalalzadeh and A. H. Ziaie, Eur. Phys. J. C {\bf 75} 53 (2015).
\bibitem{grcolapse1} A. H. Ziaie, P. V. Moniz, A. Ranjbar and H. R. Sepangi, Eur. Phys. J. C {\bf 74} (2014) 3154.
\bibitem{highertor} M. W. Kalinowski, Lett. Math. Phys. {\bf 5} 489 (1981).
\bibitem{highertor1} G. German, A. Macias and O. Obregon, Class. Quantum Grav. {\bf 10} 1045 (1993)
\bibitem{highertor2}  R. Troncoso and J. Zanelli, Class. Quant. Grav. {\bf 17} 4451 (2000);\\ A. W. Smith, Z Phys. C, {\bf 24} 85 (1984).
\bibitem{bhtorsion} S. N. Solodukhin, Phys. Lett. B, {\bf 319} 87 (1993).
\bibitem{bhtorsion1} B. Cvetkovic, M. Blagojevic Mod. Phys. Lett. A {\bf 22} 3047 (2007).
\bibitem{bhtorsion2} M.-S. Ma, F. Liu and R. Zhao, Class. Quantum Grav. {\bf 31} 095001 (2014).
\bibitem{bhtorsion3} M. Blagojevic and B. Cvetkovic, JHEP {\bf 05} 101 (2015).
\bibitem{bhtorsion4} N. J. Poplawski, Astrophys. J. {\bf 832} 96 (2016).
\bibitem{bhtorsion5} J. A. R. Cembranos and J. G. Valcarcel, JCAP {\bf 1701} 014 (2017).
\bibitem{strongfieldbh} D. Psaltis, Living Rev. Relativity, {\bf 11} 9 (2008).
\bibitem{highlagdyntor} Y. N. Obukhov, V. N. Ponomarev and V. V. Zhytnikov, Gen. Rel. Grav. {\bf 21} 1107 (1989);\\
Y. N. Obukhov, Int. J. Geom. Meth. Mod. Phys. {\bf 3} 95 (2006).
\bibitem{weyspinsource} J. Weyssenhoff, A. Raabe, Acta Phys. Pol. {\bf 9} 7 (1947).
\bibitem{weyspinsource1} Y. N. Obukhov, V.A. Korotky, Class. Quantum Gravity {\bf 4} 1633 (1987).
\bibitem{weyspinsource2} J. Weyssenhoff, in Max-Planck-Festschrift-1958, eds. B. Kockel
et al. (Deutscher Verlag  Wissenschaft, Berlin, 1958) p. 155.
\bibitem{weyspinsource3} F. Halbwachs, Theorie Relativiste des Fluides a Spin (GauthierVillars, Paris, 1960).
\bibitem{weyspinsource4} G. A. Maugin, Sur les fluides relativistes a spin. Ann. Inst. Henri Poincare {\bf 20} 41 (1974).
\bibitem{weyspinsource5} J. R. Ray, L. L. Smalley, Phys. Rev. D {\bf 27} 1383 (1983).
\bibitem{restorcomp} M. Tsamparlis, Phys. Lett. A {\bf 75} 27 (1979); Phys. Rev. D {\bf 24} 1451 (1981).
\bibitem{PB1981} P. Baekler, Phys. Lett. B {\bf 99} 329 (1981).
\bibitem{semieu} J. K. Seem, P. E. Ehrlich, and K. L. Easley, \lq\lq{}{\it Global Lorentzian Geometry}\rq\rq{}, Marcel Dekker, Inc. (1996).
\bibitem{semieu1}  B. O\rq{}Neill, \lq\lq{}{\it Semi-Riemannian Geometry, with applications to relativity,}\rq\rq{} Academic Press (1983).
\bibitem{semieu2} M. F. Atiyah and R. S. Ward, Commun. Math. Phys. {\bf 55} 117 (1977).
\bibitem{semieu3} R. S. Ward, Phil. Trans. R. Soc. London, Ser. A {\bf 315} 451 (1985).
\bibitem{semieu4} M. A. De Andrade, O. M. Del Cima and L. P. Colatto, Phys. Lett. B {\bf 370} 59 (1996).
\bibitem{semieu5} M. A. De Andrade and O. M. Del Cima, Int. J. Mod. Phys. A {\bf 11} 1367 (1996).
\bibitem{semieu6} M. Carvalho and M. W. de Oliveira, Phys. Rev. D {\bf 55} 7574 (1997).
\bibitem{SDSGSS} H. Garcia-Compean, \lq{}\lq{}{\it $N = 2$ string geometry and the heavenly equations,}\rq{}\rq{} in Proc.
Conf. on Topics in Mathematical Physics, General Relativity and Cosmology in Honor of Jerzy Plebanski, Mexico City, Mexico eds. H. Garcia Compean et al. (World Scientific, 2006), hep-th/0405197.
\bibitem{SDSGSS1} H. Ooguri and C. Vafa, Nucl. Phys. B {\bf 367} 83 (1991).
\bibitem{SDSGSS2} H. Ooguri and C. Vafa, Nucl. Phys. B {\bf 361} 469 (1991).
\bibitem{SDSGSS3} E. Bergshoeff and E. Sezgin, Phys. Lett. B {\bf 292} 87 (1992).
\bibitem{SDSGSS4} S. James Gates Jr. H. Nishino and S. V. Ketov, Phys. Lett. B {\bf 297} 99 (1992).
\bibitem{SDSGSS5} S. V. Ketov, H. Nishino and S. James Gates, Jr., Nucl. Phys. B {\bf 393} 149 (1993).
\bibitem{SDSGSS6} D. Kutasov, E. J. Martinec and M. O’Loughlin, Nucl. Phys. B {\bf 477} 675 (1996).
\bibitem{SDSGSS7} J. Martinec, \lq\lq{}{\it Matrix theory and N = (2; 1) strings,}\rq\rq{} hep-th/9706194.
\bibitem{SDSGSS8} D. Kutasov and E. J. Martinec, Nucl. Phys. B {\bf 477} 652 (1996).
\bibitem{SDSGSS9} D. Kutasov and E. J. Martinec, Class. Quantum Grav. {\bf 14} 2483 (1997).
\bibitem{spinor2+2} S. V. Ketov, H. Nishino and S. James Gates Jr., Phys. Lett. B {\bf 307} 323 (1993).
\bibitem{spinor2+21} P. G. O. Freund, \lq\lq{}{\it Introduction to Supersymmetry}\rq\rq{} (Cambridge University Press, 1986).
\bibitem{2+2cosmo} J. A. Nieto, M. P. Ryan, O. Velarde and C. M. Yee, Int. J. Mod. Phys. A {\bf 19} 2131 (2004).
\bibitem{2+2BH} C. Castro and J. A. Nieto, Int. J. Mod. Phys. A {\bf 22} 2021 (2007).
\bibitem{Joshibook} P. S. Joshi, \lq\lq{}{\it Gravitational Collapse and Space-Time Singularities}\rq\rq{}, (Cambridge: Cambridge
University Press, 2007).
\bibitem{nakedinfomedium} S. K. Chakrabarti and P. S. Joshi, Int. J. Mod. Phys. D {\bf 3} 647 (1994).
\bibitem{nakedinfomedium1} M. Patil and P. S. Joshi, Phys. Rev. D {\bf 82} 104049 (2010).
\bibitem{nakedinfomedium2} P. S. Joshi, D. Malafarina, Int. J. Mod. Phys. D {\bf 20} 2641 (2011).
\bibitem{nakedinfomedium3} M. Patil, P. S. Joshi and D. Malafarina, Phys. Rev. D {\bf 83} 064007  (2011).
\bibitem{nakedinfomedium4} M. Patil, P. S. Joshi, M. Kimura and K.-I. Nakao, Phys. Rev. D {\bf  86} 084023 (2012).
\bibitem{nakedinfomedium5} M. Patil and P. S. Joshi, Phys. Rev. D {\bf 85} 104014 (2012).
\bibitem{nakedinfomedium6} Z. Stuchlik and J. Schee, Class. Quantum Grav. {\bf 31} 195013 (2014).
\bibitem{nakedinfomedium7} R. S. S. Vieira, J. Schee, W. Kluzniak, Z. Stuchlik and M. Abramowicz, Phys. Rev. D {\bf 90} 024035 (2014).
\bibitem{nakedinfomedium8} F. S. Khoo and Y. C. Ong, Class. Quantum Grav. {\bf 33} 235002 (2016).
\bibitem{gibhawdesitter} G. W. Gibbons and S. W. Hawking, Phys. Rev. D {\bf 15} 2738 (1977).
\bibitem{Nariaisol} H. Nariai, Gen. Rel. Grav., {\bf 31} 963  (1999); Gen. Rel. Grav., {\bf 31} 951 (1999).
\bibitem{sttg} S. Capozziello and V. Faraoni, \lq\lq{}{\it Beyond Einstein Gravity, a Survey of Gravitational Theories
for Cosmology and Astrophysics,}\rq\rq{} Springer (2011).
\bibitem{obsnakedbh} P. S. Joshi, D. Malafarina and R. Narayan, Class. Quantum Grav. {\bf 28} 235018 (2011).
\bibitem{obsnakedbh1} P. S. Joshi, D. Malafarina and R. Narayan, Class. Quantum Grav. {\bf 31} 015002 (2014 ).
\bibitem{obsnakedbh2} N. Ortiz, O. Sarbach and T. Zannias, Class. Quantum Grav. {\bf 32} 247001 (2015).
\bibitem{nakedvsblhole} K. S. Virbhadra, D. Narasimha and S. M. Chitre, Astron. Astrophys. {\bf 337} 1 (1998).
\bibitem{nakedvsblhole1} K. S. Virbhadra and C. R. Keeton, Phys. Rev. D {\bf 77} 124014 (2008).
\bibitem{nakedvsblhole2} K. S. Virbhadra and G. F. R. Ellis, Phys. Rev. D {\bf 65} 103004 (2002).
\bibitem{nakedvsblhole3} K. S. Virbhadra and G. F. R. Ellis, Phys. Rev. D {\bf 62} 084003 (2000).
\bibitem{nakedvsblhole4} C.-M. Claudel, K. S. Virbhadra and G. F. R. Ellis J. Math. Phys. {\bf 42} 818 (2001).
\bibitem{Annalsspin} S. Capozziello, G. Lambiase and C. Stornaiolo, Annalen Phys. {\bf 10} 713 (2001).
\bibitem{teleparallel} C. Moller, K. Dan. Vidensk. Selsk. Mat. Fys. Skr. {\bf 1} 1 (1961).
\bibitem{teleparallel1} E. W. Mielke Ann. Phys. {\bf 219} 78 (1992).
\bibitem{teleparallel2} J. W. Maluf, J. Math. Phys. {\bf 35} 335 (1994).
\bibitem{teleparallel3} V. C. de Andrade and J. G. Pereira, Phys. Rev. D {\bf 56} 4689 (1997).
\bibitem{teleparallel4} V. C. De Andrade, L. C. T. Guillen and J. G. Pereira gr-qc/0011087.
\bibitem{fTLags} G. R. Bengochea and R. Ferraro, Phys. Rev. D {\bf 79} 124019 (2009).
\bibitem{fTLags1} E. V. Linder, Phys. Rev. D {\bf 81} 127301 (2010).
\bibitem{fTLags2} V. K. Oikonomou and E. N. Saridakis, Phys. Rev. D {\bf 94} 124005 (2016).
\bibitem{fTLags3} R. Ferraro and F. Fiorini, Phys. Rev. D {\bf 75} 084031 (2007).
\bibitem{fTLags4} R. Ferraro and F. Fiorini, Phys. Rev. D {\bf 78} 124019 (2008).
\bibitem{fTgrav} Y.-F. Cai, S. Capozziello, M. De Laurentis and E. N. Saridakis, Rep. Prog. Phys. {\bf 79} 106901 (2016).
\bibitem{extendfTgrav} H. Abedi, S. Capozziello, R. D\rq{}Agostino and O. Luongo, arXiv:1803.07171 [gr-qc].
~

\end{thebibliography}
\end{document}